\documentclass[11pt]{article}
  
\usepackage{amsmath,amssymb}

\usepackage{epsfig}  
\usepackage{graphicx}               
\usepackage{url}
\usepackage{hyperref}
\usepackage{slashed}
\usepackage[makeroom]{cancel}
\usepackage{cite}

\usepackage{pstricks}
\usepackage{color}

\usepackage{tikz}
\usetikzlibrary{positioning}
\usetikzlibrary{decorations.pathmorphing}
\usetikzlibrary{decorations.markings}
\usetikzlibrary{snakes}
 
\newcommand{\nc}{\newcommand}  

\nc{\beq}{\begin{equation}}  
\nc{\eeq}{\end{equation}}  
\nc{\beqa}{\begin{eqnarray}}  
\nc{\eeqa}{\end{eqnarray}}  
\nc{\bea}{\begin{eqnarray}}  
\nc{\eea}{\end{eqnarray}}  
\nc{\ra}{\rightarrow}  
\nc{\lsim}{\begin{array}{c}\,\sim\vspace{-21pt}\\< \end{array}}  
\nc{\gsim}{\begin{array}{c}\sim\vspace{-21pt}\\> \end{array}}  
\nc{\Tr}{{\rm Tr}}
\nc{\slsh}{\slash\hspace*{-0.22cm}}

\def\be{\begin{equation}}
\def\ee{\end{equation}}
\def\bea{\begin{eqnarray}}
\def\eea{\end{eqnarray}}
\def\bit{\begin{itemize}}
\def\eit{\end{itemize}}

\newcommand{\missET}{\slash{\hspace{-2.5mm}E}_T}

\def\to{\rightarrow}

\title{  
\vspace*{-2.3cm}  
\begin{flushright}  
\normalsize{  
SLAC-PUB-15704
  }  
\end{flushright}  
\vspace{1.5cm}  
\Large  
\textbf{
Fermion Portal Dark Matter
}\vspace*{1.0cm}   
}

\author{Yang Bai\,$^{a}$ and Joshua Berger\,$^{b}$
\vspace{5mm}
\\
$^{a}$ \normalsize\emph{Department of Physics, University of Wisconsin, Madison, WI 53706, USA}  \vspace{1mm} \\
$^{b}$ \normalsize\emph{SLAC National Accelerator Laboratory, 2575 Sand Hill Road, Menlo Park, CA 94025, USA}
}

\date{}

\begin{document}  

\tikzset{ 
  scalar/.style={dashed},
  scalar-ch/.style={dashed,postaction={decorate},decoration={markings,mark=at
      position .5 with {\arrow{>}}}},
  fermion/.style={postaction={decorate}, decoration={markings,mark=at
      position .5 with {\arrow{>}}}},
  gauge/.style={decorate, decoration={snake,segment length=0.2cm}},
  gauge-na/.style={decorate, decoration={coil,amplitude=4pt, segment
      length=5pt}}
}

\setcounter{page}{0}  
\maketitle  

\vspace*{1cm}  
\begin{abstract} 
We study a class of simplified dark matter models in which one dark
matter particle couples with a mediator and a Standard Model
fermion. In such models, collider and direct detection searches probe
complimentary regions of parameter space. For Majorana dark matter,
direct detection covers the region near mediator-dark matter
degeneracy, while colliders probe regions with a large dark matter and
mediator mass splitting. For Dirac and complex dark matter, direct
detection is effective for the entire region above the mass threshold,
but colliders provide a strong bound for dark matter lighter than a
few GeV. We also point out that dedicated searches for signatures with
two jets or a mono-jet not coming from initial state radiation, along
missing transverse energy can cover the remaining parameter space for
thermal relic dark matter.
\end{abstract}  
  
\thispagestyle{empty}  
\newpage  
  
\setcounter{page}{1}

\baselineskip18pt   

\vspace{-3cm}

\section{Introduction}
\label{sec:intro}
There is now a large body of evidence for the existence of particulate
dark matter that interacts gravitationally and makes up nearly a quarter of the energy density of the universe~\cite{Begeman:1991iy,Bradac:2006er,Bennett:2012zja,Ade:2013zuv}.  On the other hand, such dark matter does not exist in the Standard Model (SM), has not been conclusively observed to interact non-gravitationally, and has never
been detected, directly or indirectly, in a laboratory experiment.
Determining the physical properties of dark matter therefore
constitutes one of the most pressing questions in high-energy
physics.  Ideally, it would be possible to measure the non-gravitational
interactions in several ways: by directly detecting rare ambient dark matter
scattering events off a target in underground laboratories, by
detecting the products of dark matter 
self-annihilation or decay, and by measuring a momentum imbalance in
collider events.

Given that so little is known about the properties of dark matter, the
most informative framework for studying dark matter properties in a
simplified framework is as model-independent as possible.  One
powerful approach is to consider effective operators of dark matter
coupling to quarks and other SM particles~\cite{Birkedal:2004xn,
  Barger:2008qd,
  Cao:2009uw,Beltran:2010ww,Bai:2010hh,Goodman:2010yf,Goodman:2010ku,Fan:2010gt,Buckley:2011kk,Rajaraman:2011wf,Fox:2011pm,Cheung:2012gi,Aaltonen:2012jb,Fitzpatrick:2012ix,Barger:2012pf,Bai:2012xg,ATLAS-CONF-2012-147,Chae:2012bq,Dreiner:2013vla,Cornell:2013rza,CMS-PAS-EXO-12-048}. In
the absence a light mediator below around one GeV, the effective
operator approach provides an excellent description for interpreting the direct
detection experiment results because of the small exchanged momentum in
the scattering processes. A simplified description in terms of such operators neatly
encapsulates all possible interactions of dark matter with their
detectors.  On the other hand, such operators may be unsuitable or at
least not precise for the purposes of collider
studies~\cite{Shoemaker:2011vi,An:2012ue,Busoni:2013lha}. The latest
searches for dark matter at
ATLAS~\cite{ATLAS-CONF-2012-147,ATLAS-CONF-2013-073} and
CMS~\cite{CMS-PAS-EXO-12-048, CMS-PAS-EXO-13-004} have set a lower
bound on the cutoffs of the effective operators to be around one
TeV. The current constraints are stringent when compared to the limits
derived from direct detection experiments, but are still much below
the center-of-mass energy, 8 TeV, of the latest run at the LHC.  When
the parton-level collision energy is comparable to the mediator mass,
the effective theory does not apply and a more complete description is
required.  

To go beyond an effective operator approach, one could directly use a
concrete underlying model such as the Minimal Supersymmetric Standard
Model (MSSM) to study the  complementarity among different
experimental probes of dark
matter~\cite{Cahill-Rowley:2013dpa}.  However, to maintain the
model-independent paradigm of the effective operator approach, a
``Simplified Dark Matter Model" could be a better choice. In
this approach, one can introduce one or two new particles with one or two
interactions to probe the potentially complicated dark matter
sector.  For dark matter interacting with SM particles, one of the most
important things is the properties of the mediator. The majority of
existing studies in the literature have the mediator particle couple
to two dark matter particles at the same time: Higgs
portal~\cite{Shrock:1982kd,Burgess:2000yq,Patt:2006fw},  2HDM
portal~\cite{Bai:2012nv}, axion-portal~\cite{Nomura:2008ru},
gravity-mediated~\cite{Lee:2013bua},
dilaton-assisted~\cite{Bai:2009ms} and
$Z^\prime$-mediated~\cite{Cheung:2007ut,Feldman:2007wj}~\footnote{The milli-charged dark matter with only one massless gauge boson in this model contradicts quantum gravity, as shown in Ref.~\cite{Shiu:2013wxa} based on arguments in Ref.~\cite{Banks:2010zn}.}~\cite{Dudas:2009uq}.
In this paper, we study simplified dark matter models with SM
fermions as the portal particle, which we call Fermion Portal (FP)
dark matter. This class of models is well motivated and can easily
be a part of some underlying models such as
SUSY~\cite{Jungman:1995df} or extra-dimension
models~\cite{Bertone:2004pz}.  

In FP dark matter models, an SM singlet dark matter particle interacts with
quarks via a new QCD color triplet state.  There are several classes of
such models depending on the Lorentz properties of the dark matter: it
may be a Dirac fermion, Majorana fermion, complex scalar, real scalar,
or vector.  The vector case requires an additional scalar field or
additional dynamics to provide the vector boson mass and we do not
consider it here.  For the real scalar dark matter case, the 
non-relativistic interaction cross-sections including
self-annihilation and direct detection are highly suppressed and we do
not consider this case either.  Therefore, in this paper we perform detailed studies of the
Dirac fermion, Majorana fermion, and complex scalar cases.

A further specification can be made based on species of quarks to
which the dark matter couples. At renormalizable level, the simplest
interactions are to have dark matter particles couple to only
right-handed fermions. In the non-relativistic limit, the interactions
can be transformed as vector or axi-vector couplings between the dark
matter particle and the SM fermion. For either type of couplings, a
dark matter particle that couples exclusively to heavy quarks would
have a suppressed direct detection cross-section.  We concentrate
on the couplings to up and down quarks in this paper and leave the
heavy quark case for future exploration.

The six cases that are considered here constitute a set of
models for benchmarking the progress of dark matter experiments, as
well as for studying the complementarity of different types of
experiments.  Unlike the effective operator approach, they can be extrapolated
to arbitrarily high energies while giving sensible results.
Furthermore, they are well-motivated and simple enough to allow for
deep experimental scrutiny.  In this paper, we demonstrate the power 
of current experiments to probe these models and determine the allowed
parameter space to probe in the future. For each case, we also show
the parameter space to satisfy the dark matter thermal relic
abundance. We pay attention to the potential for current experiments
to probe the thermal relic hypothesis.  We find that the parameter
space for a thermal relic is highly constrained for most of the
scenarios considered, but that there is some potential at the moment
in a model with Majorana dark matter. 

The remainder of this paper is structured as follows.  In
Section~\ref{sec:quark}, we introduce the Fermion Portal class of
simplified models.  We determine the allowed parameter space for dark
matter to be a thermal relic in Section~\ref{sec:relic-abundance}.
Current direct detection and collider constraints are determined in
Sections~\ref{sec:direct-detection} and \ref{sec:collider}
respectively, with summary plots presented in
Section~\ref{sec:collider}.  We discuss potential improvement for the
LHC collider searches and conclude in Section~\ref{sec:conclusion}.

\section{Simplified dark matter model: fermion portal}
\label{sec:quark}
If the dark matter sector interacts directly with a single fermion in the SM,
two particles with different spins are required in the dark matter
sector.  In this paper, we will concentrate on the quark portal dark
matter and leave the lepton portal dark matter for future
exploration. Restricting to particles with a spin less than one, there
are two general situations: fermionic dark matter with a color-triplet
scalar partner or scalar dark matter with a color-triplet fermion
partner.  In the former case, we consider both Dirac and Majorana dark
matter, while for the latter case we only consider a complex scalar
dark matter and skip the real scalar dark matter
case~\cite{Barger:2008qd}, which has a quark mass suppressed $s$-wave
or a $d$-wave or three-body 
suppressed annihilation rate and a velocity suppressed direct
detection cross section if the quark masses are neglected.

We begin by considering fermionic dark matter coupled to right-handed quarks as the
portal to the dark matter sector.  The dark matter candidate may be a Dirac or
Majorana fermion, $\chi$, that is an SM 
gauge singlet.  The mediator is an $SU(3)_c$ triplet with an appropriately
chosen hypercharge.   The renormalizable operators are 
\beqa
{\cal L}_{\rm fermion} \supset \lambda_{u_i} \phi_{u_i} \overline{\chi}_L u^i_R \,+\,   \lambda_{d_i} \phi_{d_i} \overline{\chi}_L d^i_R \,+\, \mbox{h.c.} \,,
\label{eq:lag-fermion}
\eeqa
where $u_i = u, c, t$ ($d_i = d, s, b$) are different  SM
quarks. Since $\chi$ is the dark matter candidate, the partner masses 
$m_{\phi_i}$ must be larger than the dark matter mass $m_\chi$. In
our analysis, we assume the branching ratio of the decay $\phi_{u_i}
\rightarrow \chi \bar{u}^i$ and $\phi_{d_i} \rightarrow \chi
\bar{d}^i$ is 100\%. We also require the Yukawa couplings
$\lambda_i$ to be less than $\sqrt{4\pi}$ to preserve
perturbativity. Since we will concentrate on the first generation
quarks, we neglect the flavor index from now on to simplify the
notation. Using the up quark operator, the width of $\phi_u$ particle
is calculated to be
\beqa
\Gamma(\phi \rightarrow \chi + \overline{u}) \,=\, \frac{\lambda_{u}^2}{16\pi}\,\frac{ (m_\phi^2 - m_\chi^2)^2}{m_\phi^3} \,,
\eeqa
for both Dirac and Majorana cases.

Similarly, for a complex scalar dark matter, $X$, and its partner, $\psi$, a color-triplet Dirac fermion, we have the interactions
\beqa
{\cal L}_{\rm scalar}  \supset \lambda_{u_i} X \overline{\psi}^{u_i}_L u^i_R \,+\,  \lambda_{d_i} X \overline{\psi}^{d_i}_L d^i_R \,+\, \mbox{h.c.} \,. 
\label{eq:lag-scalar}
\eeqa
For the up quark operator, the decay width of $\psi^u$ field is
\beqa
\Gamma(\psi \rightarrow X^\dagger + u) \,=\,  \frac{\lambda_{u}^2}{32\pi}\,\frac{ (m_\psi^2 - m_X^2)^2}{m_\psi^3} \,. 
\eeqa

If the operators in Eqs.~(\ref{eq:lag-fermion}), (\ref{eq:lag-scalar})
are defined in the flavor basis, the quark right-handed currents
become physical and additional (weak) flavor constraints apply to the model
parameter space.  However, if they are defined in the quark mass
basis, there are no additional flavor changing processes beyond the
SM. We simply take the mass basis assumption and ignore the flavor
physics constraints.  We next explore the dark matter phenomenology of
this class of models, including its thermal relic abundance, direct
detection and collider searches. Some other studies for the spin-dependent direct detection and indirect detection signatures can be found in Refs.~\cite{Agrawal:2010fh,Garny:2013ama}.

\section{Relic abundance}
\label{sec:relic-abundance}
The complimentarity between dark matter collider and direct detection
searches is independent of the dark matter thermal history. Since the
weakly interacting massive particle (WIMP) is still the best motivated
scenario that generates the observed dark matter relic abundance for a
weak-scale mass, we calculate the thermal relic abundance
for the simplified fermion-portal dark matter.   We then compare the
allowed thermal relic parameter space to direct detection and collider
bounds.

In the fermionic dark matter case, the main annihilation channel is
$\overline{\chi}\chi \rightarrow \overline{u} u$ for Dirac dark
matter. The dominant contribution to the annihilation cross-section is 
\beqa
\label{eq:dirac-ann}
\frac{1}{2}\,(\sigma v)^{\chi\bar\chi}_{\rm{Dirac}}  =\frac{1}{2}\left[ \frac{3\,\lambda_u^4 m_\chi^2}{32\,\pi\, (m_\chi^2 + m_{\phi}^2)^2}
\,+\, v^2 \, \frac{\lambda_u^4\,m_\chi^2 \,( - \,5 m_\chi^4 \,-\, 18
  m_\chi^2 m_{\phi}^2 + 11 m_{\phi}^4 ) }{256\,\pi\, (m_\chi^2 +
  m_{\phi}^2)^4 } \right] \equiv s + p\,v^2  \,,
\eeqa
where $v$ is the relative velocity of two dark matter particles and is
typically $0.3\,c$ at the freeze-out temperature and $10^{-3}\,c$ at
present.   The factor of 1/2 in Eq.\ (\ref{eq:dirac-ann}) accounts for
the fact that Dirac dark matter is composed of both a particle and an
anti-particle. For Majorana dark matter, the annihilation 
rate only contains a $p$-wave contribution at leading order in the limit of zero
quark masses
\beqa
(\sigma v)^{\chi\chi}_{\rm{Majorana}} = v^2\,\frac{\lambda_u^4\,m_\chi^2\,(m_\chi^4 + m_\phi^4)}{16\pi\,(m_\chi^2 + m_\phi^2)^4} \equiv p\,v^2 \,.
\eeqa

In the non-degenerate parameter space, we only need to care about the
dark matter annihilation rate. The dark matter relic abundance is
approximately related to the  ``$s$" and ``$p$" variables by 
\beqa
\Omega_\chi h^2 \approx \frac{1.07\times 10^9}{\mbox{GeV}\,M_{\rm Pl}\,\sqrt{g^*} } \,\frac{x_F}{s + 3\,(p -s/4)/x_F} \,,
\label{eq:sigmav}
\eeqa
where the Planck scale is $M_{\rm Pl} = 1.22\times 10^{19}$~GeV and
$g^*$ is the number of  relativistic degrees of freedom at the
freeze-out temperature and is taken to be 86.25 here. The freeze-out
temperature $x_F$ is given by
\beqa
x_F = \ln{\left[ \frac{5}{4} \sqrt{\frac{45}{8} } \frac{g}{2\pi^3} \frac{M_{\rm Pl} \,m_\chi (s + 6\,p /x_F)  }{\sqrt{g^*} \sqrt{x_F} }  \right]}\,,
\label{eq:xF}
\eeqa
where $g=2 (4)$ is the number of degrees of freedom for the Majorana
(Dirac) fermion dark matter. 

In the degenerate parameter space with $\Delta \equiv (m_\phi -
m_\chi)/m_\chi \ll 1$ and comparable to or below the freeze-out
temperature $1/x_F\sim 5\%$, co-annihilation effects become
important~\cite{Griest:1990kh, Edsjo:1997bg}.  Neglecting the sub-leading
electroweak interaction, the annihilation cross-section for $ \chi +
\phi^\dagger \rightarrow u + g$ is given by
\beqa
(\sigma v)^{\chi\,\phi^\dagger} \, =\, \frac{g_s^2\,\lambda_u^2}{24\pi\,m_\phi (m_\chi + m_\phi)} \,+\, v^2\, \frac{g_s^2\,\lambda_u^2 (29 m_\chi^2 - 50 m_\chi m_\phi + 9 m_\phi^2) }{576\pi\,m_\phi (m_\chi + m_\phi)^3 } \,,
\eeqa
for both Dirac and Majorana dark matter. Additionally, the $\phi$
field self-annihilation cross-section is given by
\beqa
(\sigma v)^{\phi\,\phi^\dagger}[gg] &=& \frac{7\,g_s^4}{216\pi\,m_\phi^2} \,-\, v^2\,\frac{59\,g_s^4}{5184\pi\,m_\phi^2} \,,\\
(\sigma v)^{\phi\,\phi^\dagger}[f\bar{f}] &=& v^2\frac{g_s^4}{432\pi\,m_\phi^2} \,,   \qquad \mbox{for  } f \neq u \,,\\
(\sigma v)^{\phi\,\phi^\dagger}[u\bar{u}] &=& v^2\left[ \frac{g_s^4}{432\pi\,m_\phi^2}  -  \frac{g_s^2\,\lambda_u^2}{108\pi\,(m_\chi^2 + m_\phi^2)} + \frac{\lambda_u^4\,m_\phi^2}{48\pi\,(m_\chi^2 + m_\phi^2)^2 }  \right] \,,
\eeqa
for both Dirac and Majorana dark matter. Here, $f$ represents the SM quarks and we have neglected all quark masses in our calculation for a heavy $m_\phi$ with $m_\phi \gg
m_f$. For Majorana dark matter, there is an additional annihilation channel with cross section
\beqa
(\sigma v)^{\phi\,\phi}[uu] &=& \frac{\lambda_u^4\,m_\chi^2}{6\pi\,(m_\phi^2 + m_\chi^2)^2} \,+\, v^2\,\frac{\lambda_u^4\,m_\chi^2 (3 m_\chi^4 - 18 m_\chi^2 m_\phi^2 -m_\phi^4) }{144\pi\,(m_\phi^2 + m_\chi^2)^4} \,. 
\eeqa
Following Refs.~\cite{Griest:1990kh, Edsjo:1997bg}, we have the effective
degrees of freedom as a function of the temperature parameter $x$ 
\beqa
g_{\rm eff} = g_\chi + g_\phi (1+ \Delta)^{3/2} e^{-x\,\Delta} \,,
\eeqa
with $g_\phi = 6$ (we count $\phi$ and $\phi^\dagger$ together) and $g_\chi =2 (4)$ for Majorana (Dirac)
fermion. The effective annihilation cross section for the Dirac case is
\beqa
(\sigma v)_{\rm eff} \,=\,\frac{1}{2}\, (\sigma v)^{\chi\bar\chi} \frac{g_\chi^2}{g^2_{\rm eff} }  + (\sigma v)^{\chi\phi^\dagger} \frac{g_\chi g_\phi}{g^2_{\rm eff} } (1+\Delta)^{3/2} e^{-x\,\Delta} + \frac{1}{2}\,(\sigma v)^{\phi \phi^\dagger} \frac{g_\phi^2}{g^2_{\rm eff} } (1+\Delta)^{3} e^{-2\,x\,\Delta} \,, 
\eeqa
and for the Majorana case is
\beqa
(\sigma v)_{\rm eff} \,=\, (\sigma v)^{\chi \chi} \frac{g_\chi^2}{g^2_{\rm eff} }  + (\sigma v)^{\chi\phi^\dagger} \frac{g_\chi g_\phi}{g^2_{\rm eff} } (1+\Delta)^{3/2} e^{-x\,\Delta} + \frac{1}{2}\,[(\sigma v)^{\phi \phi^\dagger} + (\sigma v)^{\phi \phi} ]\frac{g_\phi^2}{g^2_{\rm eff} } (1+\Delta)^{3} e^{-2\,x\,\Delta} \,, 
\eeqa
Variables $s_{\rm eff}$ and $p_{\rm eff}$ can be constructed 
by forming a similar combination to $(\sigma v)_{\rm eff}$.  They
replace $s$ and $p$ in Eqs.~(\ref{eq:sigmav}) and (\ref{eq:xF}) for
the purposes of calculating the thermal relic abundance.

\begin{figure}[th!]
\begin{center}
\hspace*{-0.75cm}
\includegraphics[width=0.44\textwidth]{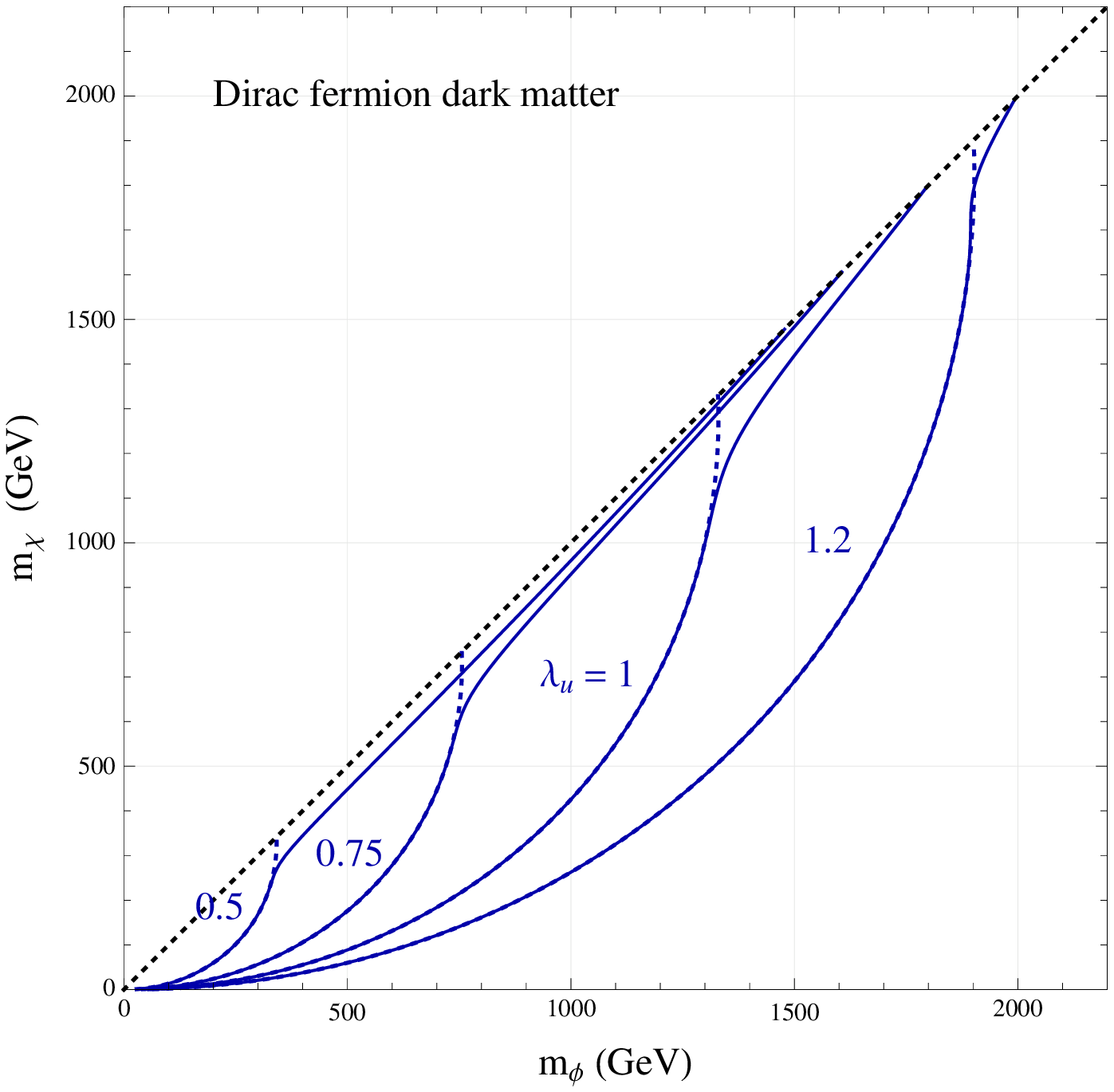} \hspace{3mm}
\includegraphics[width=0.45\textwidth]{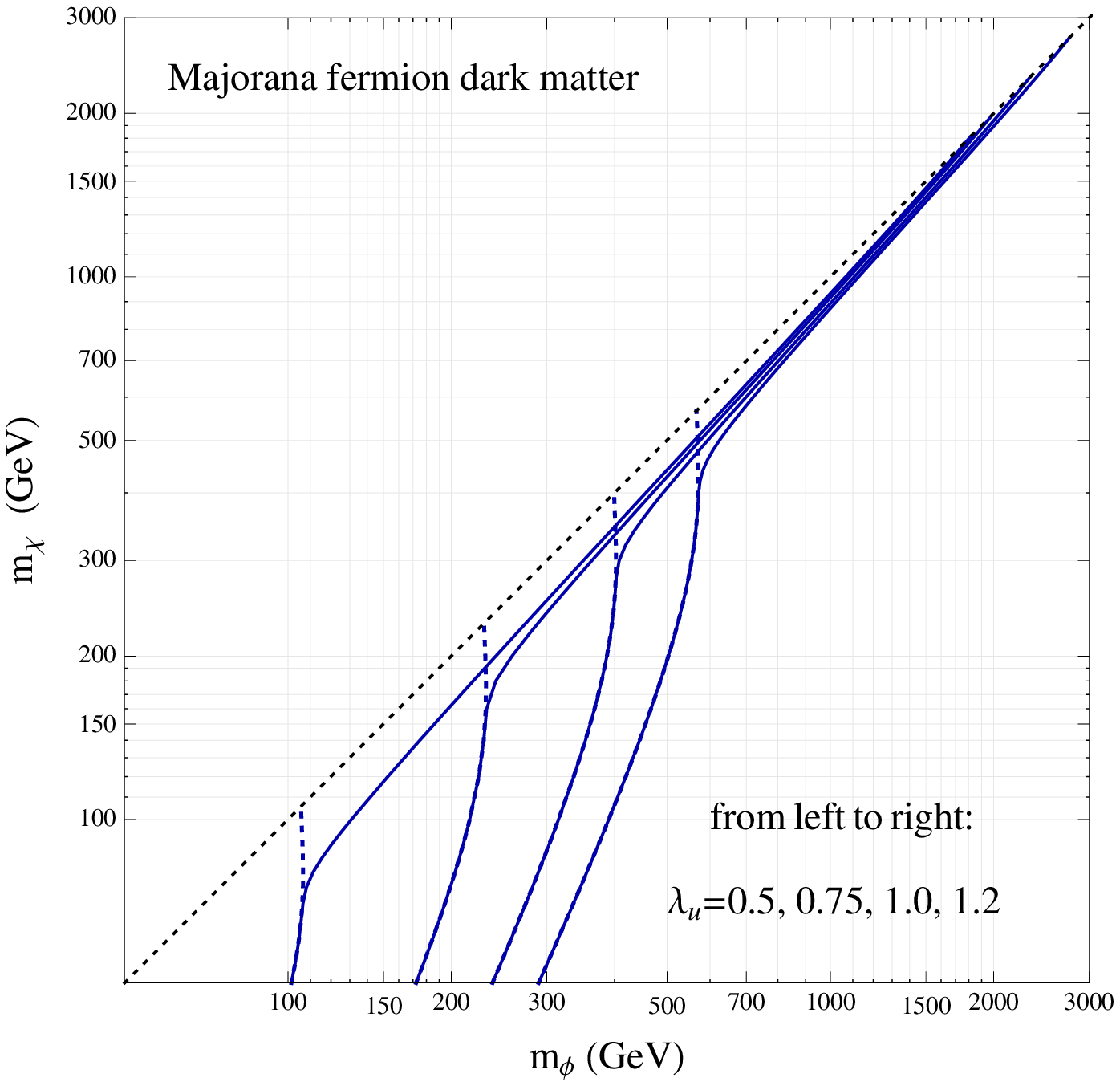}
\caption{Left panel: the masses of Dirac fermion dark matter and its
  partner for different choices of coupling, after fitting the
  observed dark matter energy fraction, $\Omega_\chi
  h^2=0.1199\pm0.0027$, from Planck~\cite{Ade:2013zuv} and
  WMAP~\cite{Bennett:2012zja}. The blue dotted lines neglect
  co-annihilation effects, while the blue solid lines include
  them. The black dotted line is boundary of the region for which $m_\phi >
  m_\chi$. Right panel: the same, but for a Majorana dark matter.}
\label{fig:relic}
\end{center}
\end{figure}

Fitting to the observed value of $\Omega_\chi h^2=0.1199\pm0.0027$
from Planck~\cite{Ade:2013zuv} and WMAP~\cite{Bennett:2012zja}, we
show the allowed values of $m_\chi$ and $m_\phi$ in
Fig.~\ref{fig:relic} for different values of couplings.  
For Dirac dark matter, the co-annihilation effects have a significant
effect for small values of $\lambda_u$, but only have a small effect
for lager values of $\lambda_u$.  Due to $p$-wave suppression of
$\chi\chi$ annihilation, Majorana dark matter mass is preferred to
have either a light mass, below around 600 GeV, or a heavy mass nearly
degenerate with its partner.

For complex scalar dark matter, the annihilation rate of $XX^\dagger
\rightarrow u\overline{u}$ is also $p$-wave suppressed and given by
\beqa
\frac{1}{2}\,(\sigma v)^{XX^\dagger}_{\rm complex\, scalar} \,=\, \frac{1}{2}\,\left[v^2\,\frac{\lambda^4\, m_X^2}{16\,\pi\,(m_X^2 + m_\psi^2)^2} \right] \equiv p\,v^2\,.
\eeqa
The allowed parameter space for a thermal relic in the
complex scalar case has similar features to the Majorana case,
including the co-annihilation effects.

\section{Dark matter direct detection}
\label{sec:direct-detection}
For calculation of dark matter direct detection cross-sections, one
could integrate out the dark 
matter partner and calculate the scattering cross sections using the
effective operators. However, for the degenerate region, the dark
matter partner in the $s$-channel can dramatically increase the
scattering cross section. To capture the resonance effects, we keep
the dark matter partner propagator in our calculation.  

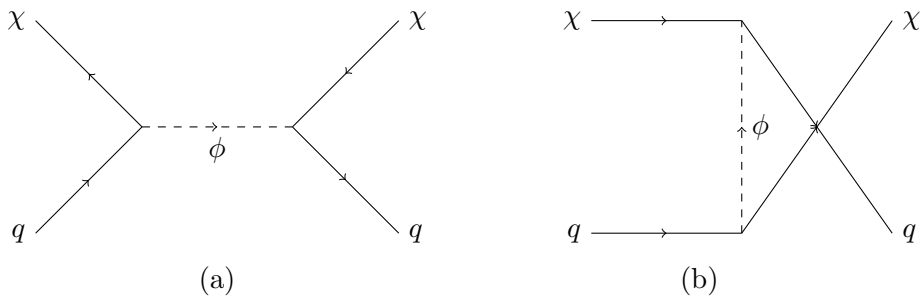
\begin{figure}[th!]
\begin{center}
\hspace*{-0.75cm}
\begin{tabular}{c c}
\begin{tikzpicture}
  \draw[fermion] (1.414,-1.414) -- (0,0) node [left] {$\chi$};
  \draw[fermion] (0,-2.828) node[left] {$q$} --  (1.414,-1.414);
  \draw[scalar-ch] (1.414,-1.414) -- node[below] {$\phi$} (3.414,-1.414);
  \draw[fermion] (4.828,0) node[right] {$\chi$} -- (3.414,-1.414);
  \draw[fermion] (3.414,-1.414) -- (4.828,-2.828) node [right] {$q$};
\end{tikzpicture} & \hspace{1cm}
\begin{tikzpicture}
  \draw[fermion]  (0,0) node [left] {$\chi$}  -- (2,0);
  \draw[fermion] (0,-2.828) node[left] {$q$} -- (2,-2.828);
  \draw[scalar-ch] (2,-2.828) -- node[right] {$\phi$} (2,0);
  \draw[fermion] (2,-2.828) node[right] {} -- (4,0) node[right] {$\chi$};
  \draw[fermion] (2,0) -- (4,-2.828) node [right] {$q$};
\end{tikzpicture} \\
(a) & (b) \\
\end{tabular}
\caption{Feynman diagrams for scattering of a fermion dark matter off nucleus. Only the left panel in (a) contributes to the Dirac fermion case, while both (a) and (b) contribute to the Majorana fermion case.}
\label{fig:dmdd}
\end{center}
\end{figure}

For the Dirac dark matter case, only the left panel in
Fig.~\ref{fig:dmdd} contributes. Both spin-independent (SI) and
spin-dependent (SD) scattering exist. The leading SI interaction
cross-section per nucleon is given by 
 \beqa
 \sigma^{Nq}_{\rm SI}({\rm Dirac}) = \frac{|\lambda_u|^4\,f^2_{N q}\, \mu^2}{64\,\pi [ (m_\chi^2 - m_\phi^2)^2 +\Gamma_\phi^2 m_\phi^2] } \,,
 \eeqa
 where $N=p, n$; $\mu$ is the reduced mass of the dark matter-nucleon system;
 $f_{N q}$ is the coefficient related to the quark operator matrix
 element inside a nucleon. For the up quark operator at hand, one has
 $f_{p\,u}=2$ and
 $f_{n\,u}=1$~\cite{Jungman:1995df,Gondolo:2004sc}. The sub-leading SD
 interaction cross section is given by
\beqa
  \sigma^{Nq}_{\rm SD} ({\rm Dirac}) =  \frac{3\,|\lambda_u|^4\,\Delta^2_{N q}\, \mu^2}{64\,\pi [ (m_\chi^2 - m_\phi^2)^2 +\Gamma_\phi^2 m_\phi^2] }\,,   \quad
    \sigma^{Nq}_{\rm SD} ({\rm Majorana}) =  \frac{3\,|\lambda_u|^4\,\Delta^2_{N q}\, \mu^2}{16\,\pi [ (m_\chi^2 - m_\phi^2)^2 +\Gamma_\phi^2 m_\phi^2] }\,,
\eeqa
with $\Delta^p_u = \Delta^n_d = 0.842\pm 0.012$ and $\Delta^p_d =
\Delta^n_u = -0.427\pm 0.013$~\cite{Belanger:2008sj}. For Majorana
dark matter, there is only an SD scattering cross section with a similar formula as the SD scattering of the Dirac fermion case.

For the complex scalar case, the SI scattering cross section is given by
 \beqa
 \sigma^{Nq}_{\rm SI}({\rm complex\; scalar}) = \frac{|\lambda_u|^4\,f^2_{N q}\, m_p^2}{32\,\pi [ (m_X^2 - m_\psi^2)^2 +\Gamma_\psi^2 m_\psi^2] } \,,
 \label{eq:si-complex-scalar}
 \eeqa
while the SD scattering cross section is suppressed by the dark matter
velocity and is neglected here.

Searches for SI dark matter interactions with
nuclei are particularly constraining when they are predicted by a
given model. We include the most stringent SI direct detection constraints from
Xenon100~\cite{Aprile:2012nq} for heavier dark matter masses and
Xenon10~\cite{Angle:2011th} for lighter dark matter masses. Xenon100 \cite{Aprile:2012nq} is sensitive to
cross-sections nearly down to $10^{-45}~{\rm cm}^2$ at a dark matter
mass of around 100 GeV.  The Xenon10\cite{Angle:2011th}
experiment has some additional sensitivity for low dark matter masses.  For the
SD scattering cross section, we mainly use the limits from SIMPLE~\cite{Felizardo:2011uw},
COUPP~\cite{Behnke:2010xt}, and PICASSO~\cite{Archambault:2009sm}
experiments for coupling to protons and from
Xenon100~\cite{Aprile:2013doa} and CDMS~\cite{Ahmed:2008eu,Ahmed:2010wy} for
coupling to neutrons.

In addition to placing strong constraints, four experiments
(DAMA\cite{Bernabei:2013cfa}, CoGeNT\cite{Aalseth:2010vx},
CRESST-II\cite{Angloher:2011uu}, CDMS\cite{Agnese:2013rvf}) have now
seen excesses in regions of parameter space already probed by
Xenon100 under some assumptions \cite{Feng:2013vod}.  For the purposes
of this study, we ignore the debatable excesses and only consider the
constraints from experiments.  Because we study the up quark and down
quark operators separately, isospin symmetry is generically
broken. Therefore, we will consider the constraints on dark
matter--proton and dark matter--neutron scattering separately for both
SI and SD.

\section{Collider constraints}
\label{sec:collider}
\begin{figure}[th!]
\begin{center}
\hspace*{-0.75cm}
\begin{tabular}{c c c}
\includegraphics[width=0.32\textwidth]{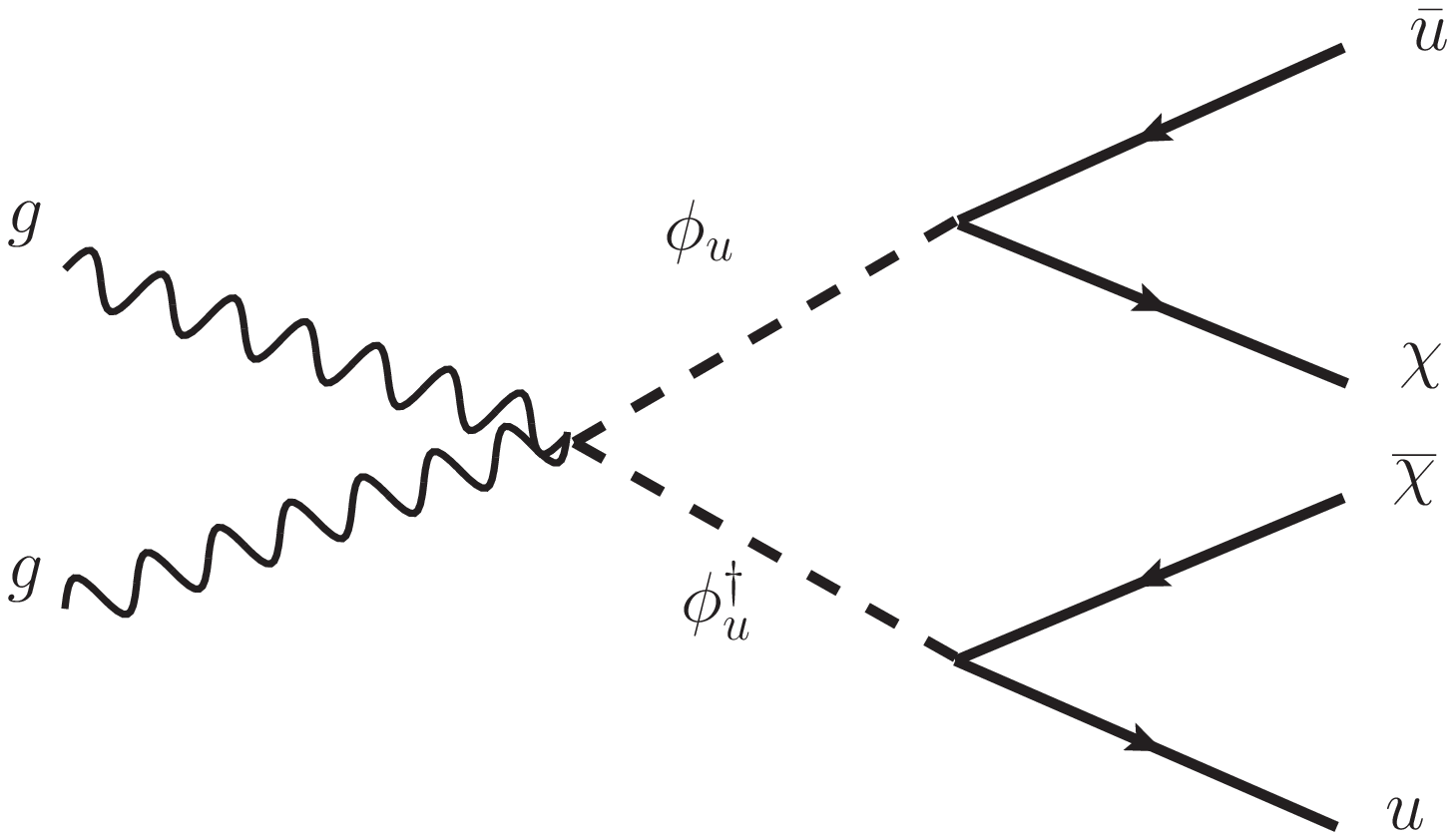} &
\includegraphics[width=0.40\textwidth]{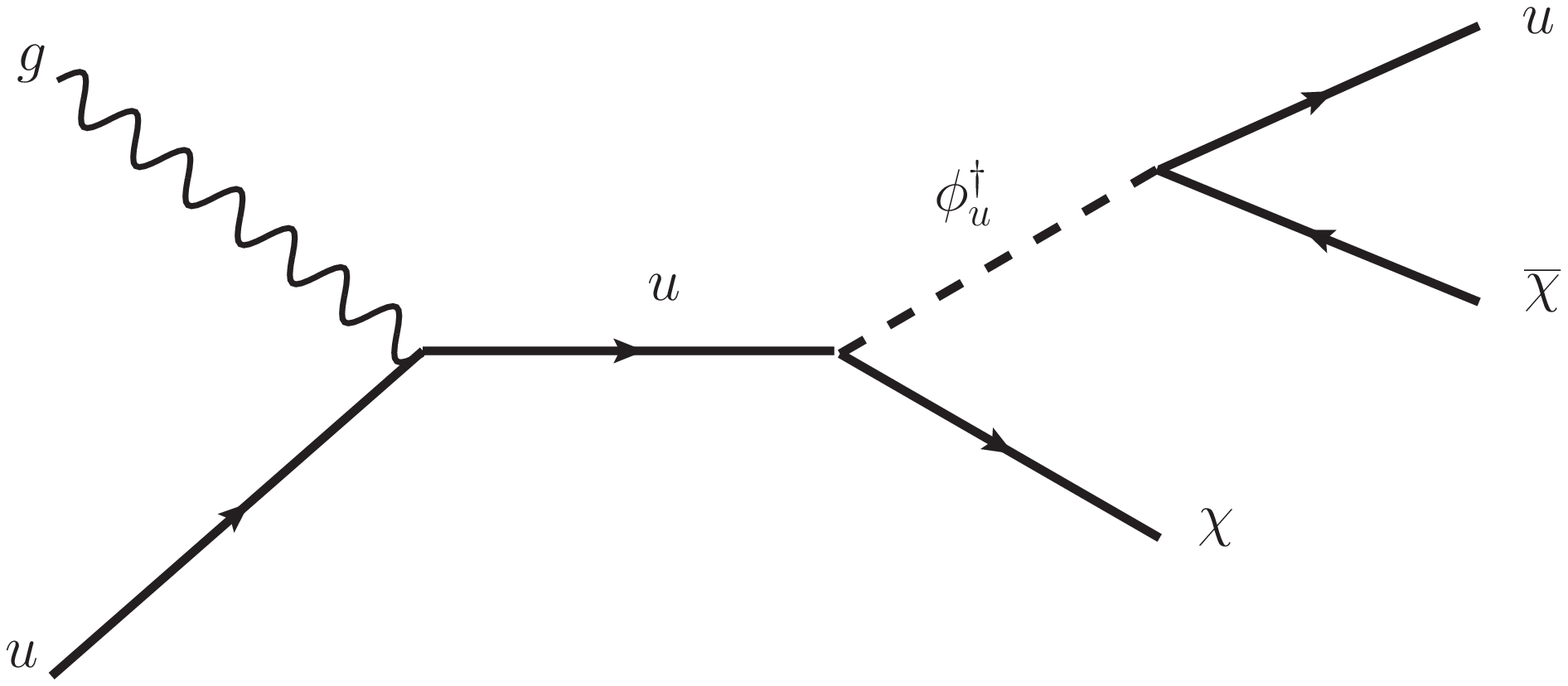} &
\includegraphics[width=0.28\textwidth]{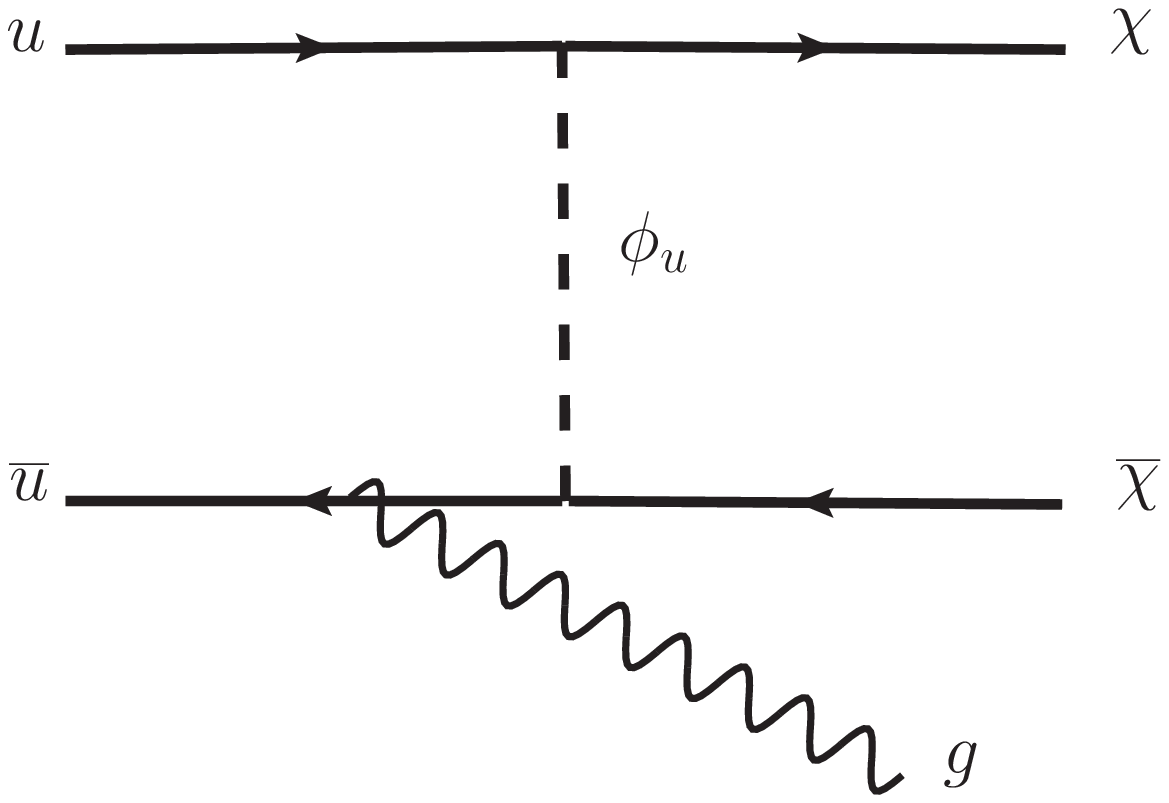} \\
(a) & (b) & (c)
\end{tabular}
\caption{The three dark matter particle production mechanisms at
  hadron colliders.  Diagram (a) has two jets in final state, while
  (b) and (c) provide mono-jet signatures.}
\label{fig:feyn1}
\end{center}
\end{figure}

Since the dark matter couples to quarks, it can be produced at
colliders.  In addition, the colored mediator may be produced,
yielding strong constraints both from associated and pair production.
Except in the regime of an extremely heavy mediator, these channels
provide the dominant constraints.  Associated production of the
mediator and the dark matter particle, along with radiative contributions from dark
matter pair production, yield a monojet signature, while pair 
production of mediators can be seen in searches for jets plus missing transverse
energy (MET).  Example diagrams for the three production
mechanisms in the case of coupling to up quarks are illustrated in
Fig.~\ref{fig:feyn1}.  For simplicity, we consider the constraints
from the CMS experiment in the monojet and jets plus MET
channels~\cite{CMS-PAS-EXO-12-048,CMS-PAS-SUS-13-012}.  There are 
comparable constraints from the ATLAS
experiment~\cite{ATLAS-CONF-2012-147,ATLAS-CONF-2013-047}, though the
reach of the CMS searches is slightly better at present.
%

\subsection{Estimated limits from monojet on $t$-change $\phi$ exchange}
\label{sec:limit-monojet-2}
For the fermionic dark matter case and in the heavy $m_{\phi}$ limit, the Fierz-transformed effective operator
\beqa
\frac{|\lambda_u|^2}{8\,m^2_{\phi}}\,\overline{\chi}\gamma_\mu\,(1+\gamma_5) \, \chi\,\overline{u}\gamma^\mu\, (1-\gamma_5) \, u \,
\eeqa
is generated.  The existing search at the 8 TeV LHC with around 20 fb$^{-1}$
constrains the combination of up quark and down quark operators. For
light dark matter masses below analysis cuts on monojet $p_T$ or
$\missET$,  the collider production cross section is insensitive to
the parity structure of the operators~\cite{CMS-PAS-EXO-12-048}. One
can approximately translate the constraints on $\Lambda \sim
\sqrt{2}\,m_{\phi}/|\lambda_u|$ obtained in
Ref.~\cite{CMS-PAS-EXO-12-048} to our model parameter space. For light
dark matter masses, the 90\% confidence level (CL) constraints on $\Lambda$ in
Ref.~\cite{CMS-PAS-EXO-12-048} is around 900 GeV, leading to an
estimated constraint of $m_{\phi}/|\lambda_u| \gtrsim 640$~GeV. 

\subsection{Limits from $2j+E_T^{\rm miss}$ on $\phi$ pair production}
\label{sec:limit-two-jet}
In the limit of a small dark matter-mediator coupling, $\lambda_u \approx 0$, the
only significant diagram yielding this final state is (a) in
Fig.~\ref{fig:feyn1}.  The production cross-section is identical to
that of a single squark in the MSSM.  The present
bounds on this process from CMS constrain the colored particle mass to
be above around 500 GeV~\cite{Chatrchyan:2013lya} for a massless
neutralino. For $\lambda_u \neq 0$, there are additional contributions from
$t$-channel dark matter exchange and  the cross-section for the 
parton level process $u + \bar{u} \to \phi + \phi^*$ is given by:
\begin{multline}
  \label{eq:pairprod-xsec}
  \sigma = - \frac{1}{1728 \pi s^3} \left\{2 \sqrt{s (s - 4 m_\phi^2)}
    \left[4 g_s^4 (4 m_\phi^2 -s) + 12 g_s^2 \lambda_u^2 (s + 2 m_\chi^2 - 2
      m_\phi^2) + 27 \lambda_u^4 s\right]  \right.\\\left.+ 3 \lambda_u^2 \left[16 g_s^2
      \left(m_\chi^2 s + (m_\phi^2 - m_\chi^2)^2\right) + 9 \lambda_u^2 s (s + 2
      m_\chi^2 - 2 m_\phi^2) \right] \log\left[\frac{s - \sqrt{s (s -4
        m_\phi^2)} + 2 m_\chi^2 - 2 m_\phi^2}{s + \sqrt{s (s -4
        m_\phi^2)} + 2 m_\chi^2 - 2 m_\phi^2} \right]\right\} \,.
\end{multline}

\begin{figure}[th!]
\begin{center}
\hspace*{-0.75cm}
\includegraphics[width=0.5\textwidth]{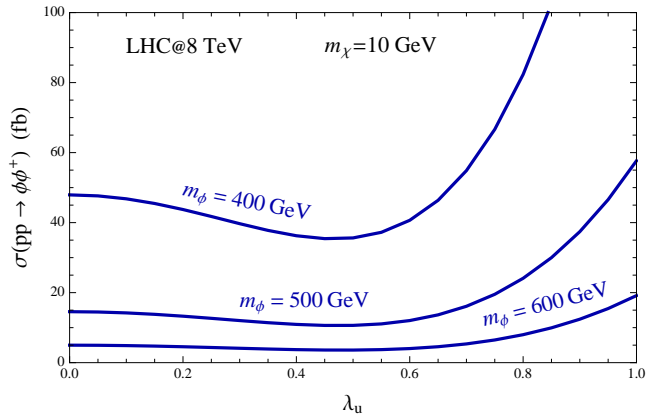}
\caption{The pair-production cross sections of the $\phi$ field as a function of $\lambda_u$.}
\label{fig:phiprod}
\end{center}
\end{figure}

 This extra contribution is
significant for $\lambda_u = 1$ and leads to a much higher
sensitivity. We also note that there is destructive interference for a
small value of $\lambda_u$, as shown in Fig.~\ref{fig:phiprod} for
different values of $m_\phi$.  We therefore anticipate that the
experimental limits from jets plus $E_T^{\rm miss}$ could become
weaker at some intermediate values of $\lambda_u$. 

To estimate the current bounds on this model, as well as the case of 
scalar dark matter, we calculate LO cross-sections for the full
process using \texttt{MadGraph} \cite{Alwall:2011uj} with a model constructed by
\texttt{FeynRules} \cite{Christensen:2008py}.  NLO K-factors calculated
using \texttt{Prospino} \cite{Beenakker:1996ed} are applied to the pure QCD contribution
to the cross-section for the cases of fermionic dark matter.  The
limits provided in \cite{CMS-PAS-SUS-13-012} are then applied to the
calculated cross-section to obtain an estimate of the current 95\% CL
exclusion limit.  The results of this analysis are presented below, in
Section \ref{sec:limit-monojet-1}.

\subsection{Limits from monojet on single $\phi$ productions}
\label{sec:limit-monojet-1}
The dominant production channel for monojets is process (b) in
Fig.~\ref{fig:feyn1} at a small value of $\lambda_u$.  The resulting cross-section at LO for $u + g \to \phi +
\chi$ is given by
\begin{equation}
  \label{eq:1}
  \sigma(u + g \to \phi + \chi) = \frac{\lambda_u^2 \,g_s^2}{768\, \pi \,s^3}
  (3 s + 2 m_\chi^2 - 2 m_\phi^2) \sqrt{(s + m_\chi^2 - m_\phi^2)^2 -
    4 m_\chi^2 s}\,,
\end{equation}
where $\sqrt{s}$ is the center-of-mass energy. 
In order to estimate the current reach of monojet searches, we
generate events for all tree-level diagrams with one
quark plus dark matter particles in the final state using \texttt{MadGraph}
\cite{Alwall:2011uj} with the models defined in \texttt{FeynRules}
\cite{Christensen:2008py}.  The events are showered and hadronized
using \texttt{Pythia} \cite{Sjostrand:2007gs}, then the hadrons are
clustered into jets using \texttt{FastJet} \cite{Cacciari:2011ma}.
The cuts described in Ref.\ \cite{CMS-PAS-EXO-12-048} are then applied
to the events in order to estimate the acceptance times efficiency of
that search.  The resulting LO signal cross section times estimated
efficiency and acceptance for each signal region are compared to the
limits set in Ref.\ \cite{CMS-PAS-EXO-12-048}.  
We present our results for several different scenarios in two ways:
first in the $m_\phi$--$m_\chi$ plane and second in the
$m_\chi$--$\sigma_{\rm SI (SD)}$ plane with all limits at $95\%$ CL.

\begin{figure}[th!]
\begin{center}
\hspace*{-0.75cm}
\includegraphics[width=0.43\textwidth]{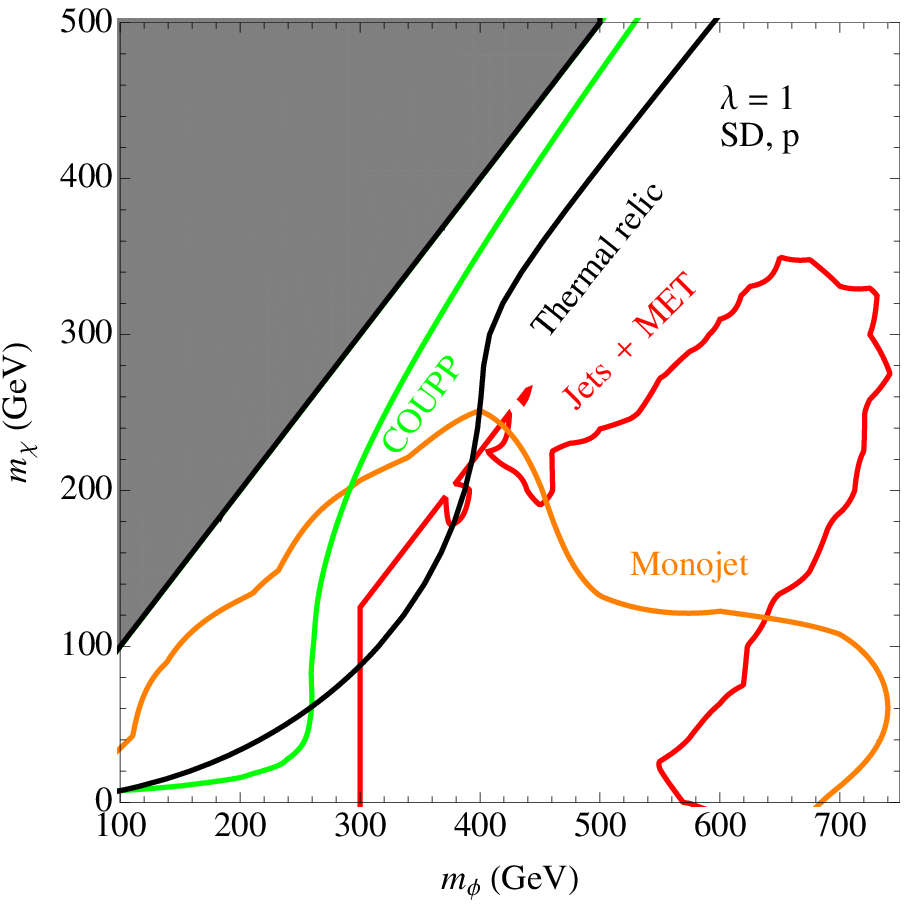} \hspace{3mm}
\includegraphics[width=0.45\textwidth]{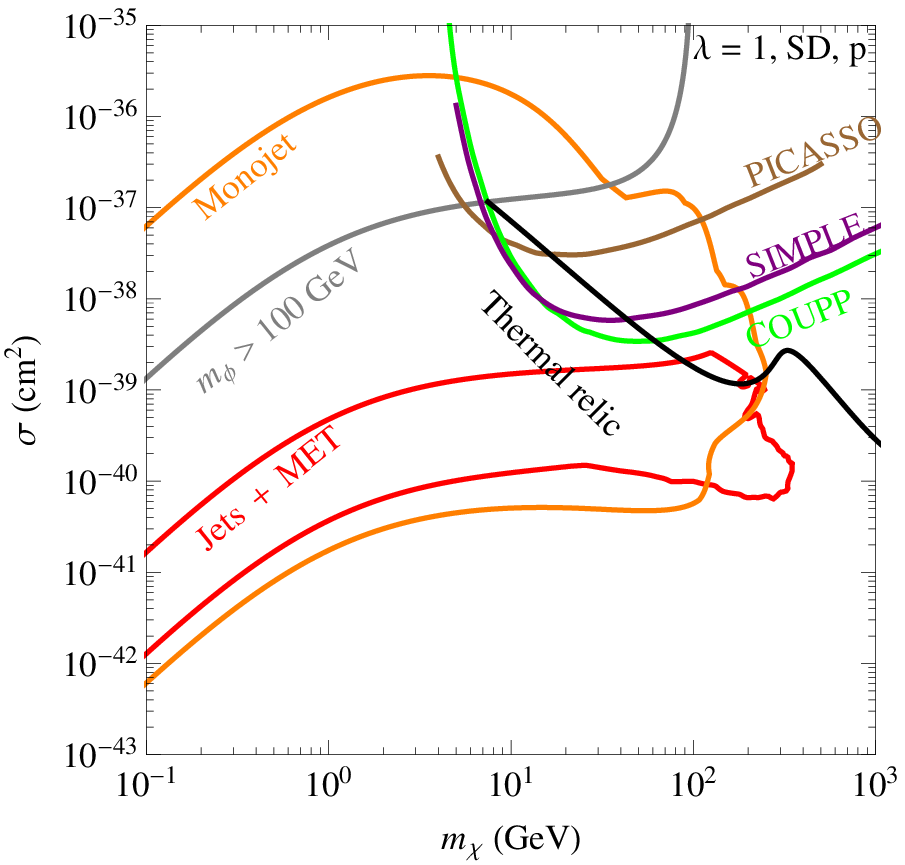}
\caption{95\% exclusion limits (except the black solid line from the thermal relic abundance) from the most sensitive searches for Majorana dark matter with the only coupling to the up quark with $\lambda_u=1$. The left panel is in the $m_\phi - m_\chi$ plane, while the right panel is in the $\sigma-m_\chi$ plane.}
\label{fig:bounds-maj-up-l1}
\end{center}
\end{figure}

We begin by considering the model with Majorana dark matter and
only $\lambda_{u} \neq 0$.  For $\lambda_{u} = 1$, the
exclusion curves are shown in Fig.~\ref{fig:bounds-maj-up-l1}.  The dominant
constraints come from collider searches in the monojet and jets + MET channels,
as well as dark matter spin-dependent direct detection searches.  In
addition, we show the lines at which the observed dark matter relic abundance
is attained assuming that $\chi$ is a thermal relic.  
The exclusion extends up to scalar masses of around $700~{\rm GeV}$ provided that
the dark matter is lighter than about $300~{\rm GeV}$.  
In Fig.~\ref{fig:bounds-maj-up-l1}, we have included the co-annihilation effects for the degenerate spectrum. 
We show the thermal relic required parameter space in the black and solid line in both panels of
Fig.~\ref{fig:bounds-maj-up-l1}. In the $\sigma-m_\chi$ plane, we stop
plotting the thermal relic line when the dark matter mass is close to
the mediator mass.  There is some parameter space at the moment where
a thermal relic is allowed, for a mediator mass of around $400~{\rm
  GeV}$, though we stress that the thermal relic abundance may be set
in other ways.  It is important to note that in 
this model, the monojet search has a wider reach than the jets + MET
search for heavy mediator masses.  This is due to the fact that some
of the diagrams for $\phi \phi$ production are proportional to the
Majorana dark matter mass.  In addition, up to dark matter masses of around
$300~{\rm GeV}$, the dominant constraint on these models comes from
colliders.  In particular, this means that the possibility of light dark
matter below a few GeV is highly constrained. The SD direct detection,
jets+MET and monojet are complimentary as they cover different parts
of parameter space. 

\begin{figure}[th!]
\begin{center}
\hspace*{-0.75cm}
\includegraphics[width=0.435\textwidth]{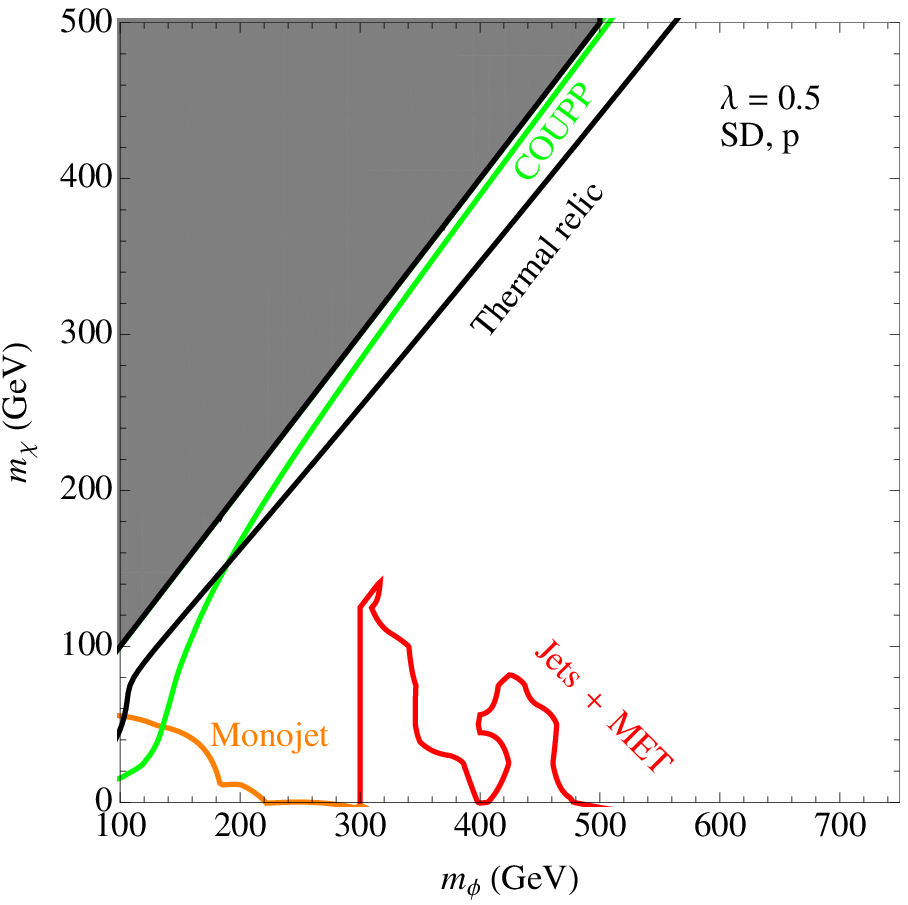}  \hspace{3mm}
\includegraphics[width=0.45\textwidth]{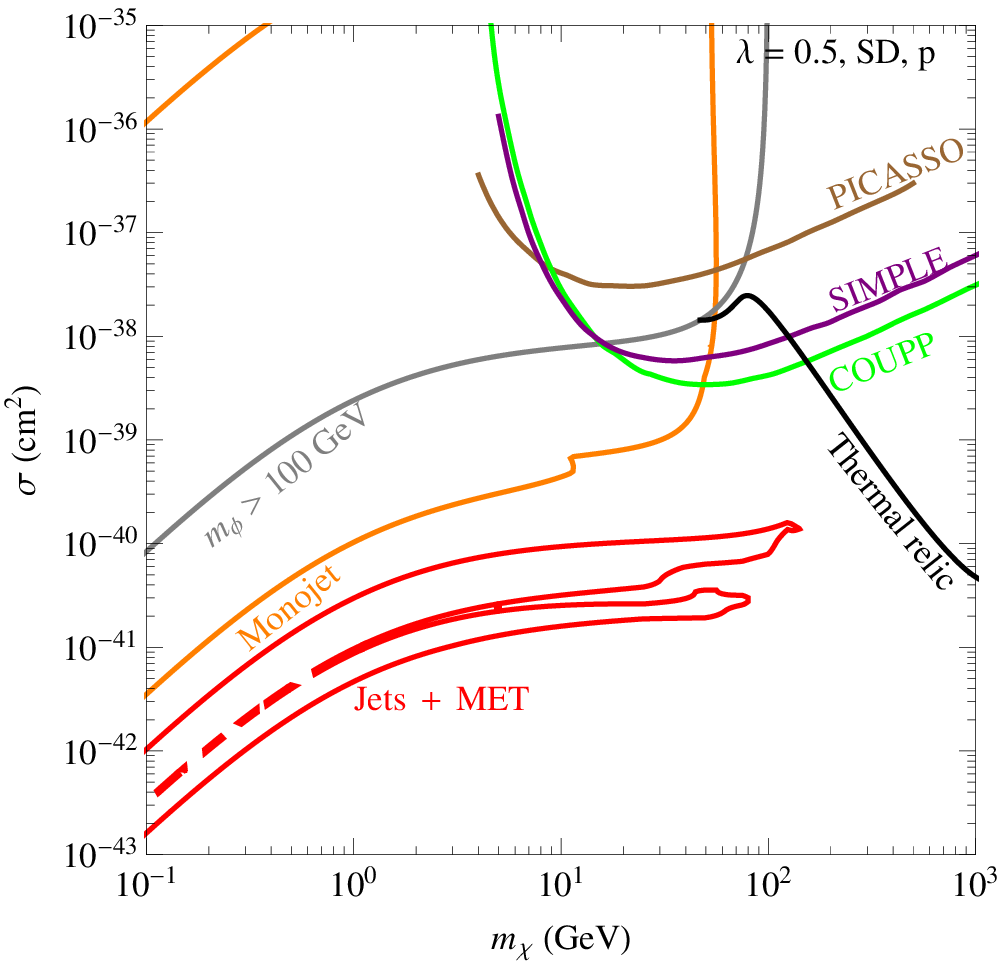}
\caption{The same as Fig.~\ref{fig:bounds-maj-up-l1} for the up quark case with $\lambda_u=0.5$.}
\label{fig:other-bounds-maj-up}
\end{center}
\end{figure}

For comparison, in Fig.~\ref{fig:other-bounds-maj-up} we show the same
exclusions in the mass plane for $\lambda_{u} = 0.5$.  In this case,
the current constraints are far weaker.  Even for the mediator masses below
a few hundred GeV, there is a significant allowed fraction of parameter space,
which it is important to cover in future searches, especially at colliders.  On the other hand,
for such a small coupling, it is difficult to obtain the correct relic
abundance via thermal production except in the co-annihilation region;
an alternate non-thermal mechanism could be considered such that dark
matter is not over-produced. 

\begin{figure}[th!]
\begin{center}
\hspace*{-0.75cm}
\includegraphics[width=0.435\textwidth]{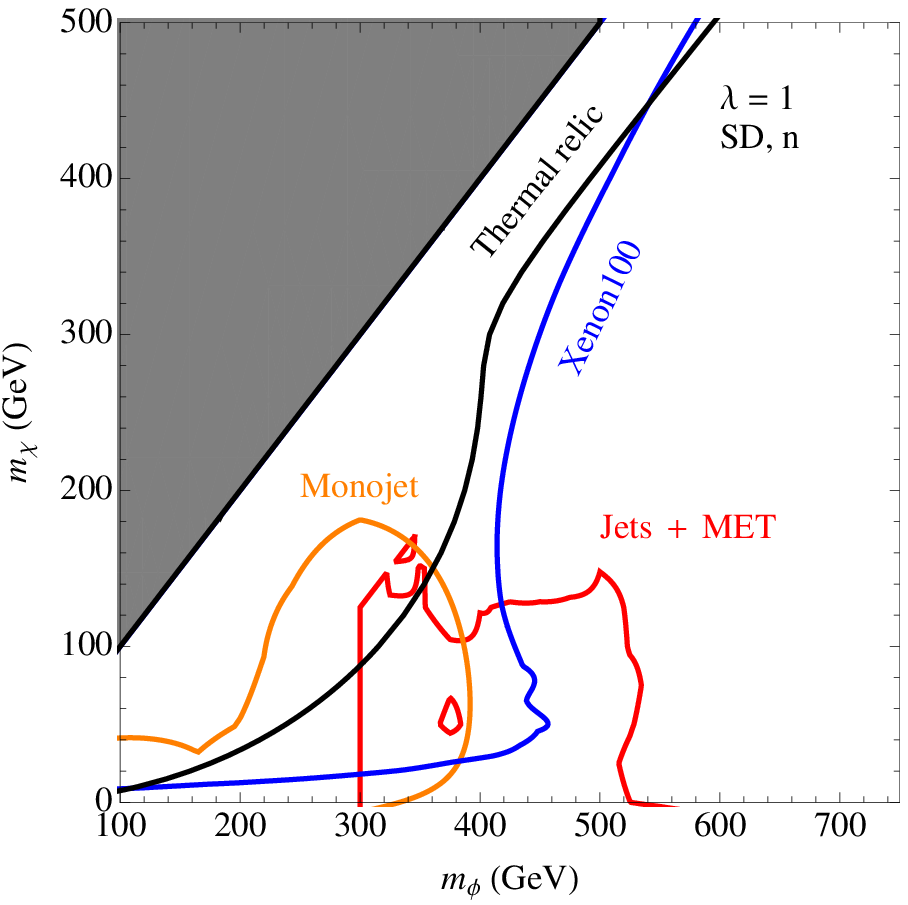} \hspace{3mm}
\includegraphics[width=0.45\textwidth]{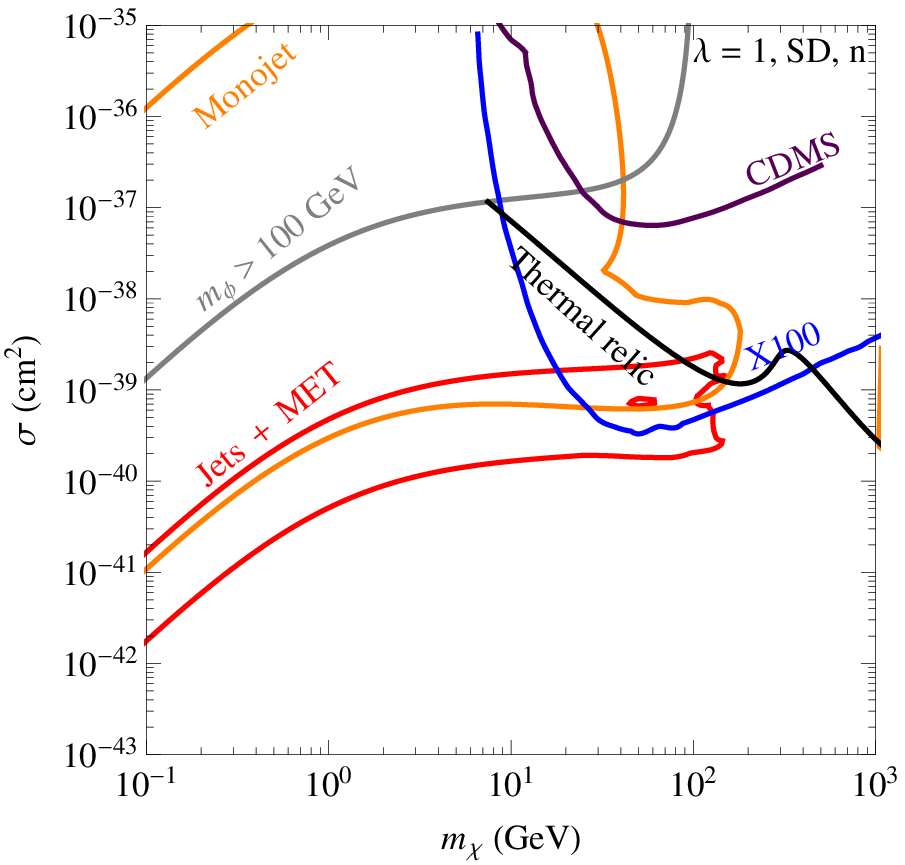}
\caption{95\% exclusion limits from the most sensitive searches for Majorana
  dark matter with coupling to the down quark.}
\label{fig:bounds-maj-down-l1}
\end{center}
\end{figure}

We also study the same model, but for the down quark case with only $\lambda_{d} \neq
0$.  For $\lambda_{d} = 1$, the exclusion curves are shown in
Figs.~\ref{fig:bounds-maj-down-l1}.  The dominant constraints are the same as in
the up-type case.  The constraints are slightly weaker in this case and
the jets + MET search dominates for at high mediator masses as it is
less sensitive to the down quark parton distribution function suppression.  In this case, there
is a similar parameter space allowed for a thermal relic.

\begin{figure}[th!]
\begin{center}
\hspace*{-0.75cm}
\includegraphics[width=0.45\textwidth]{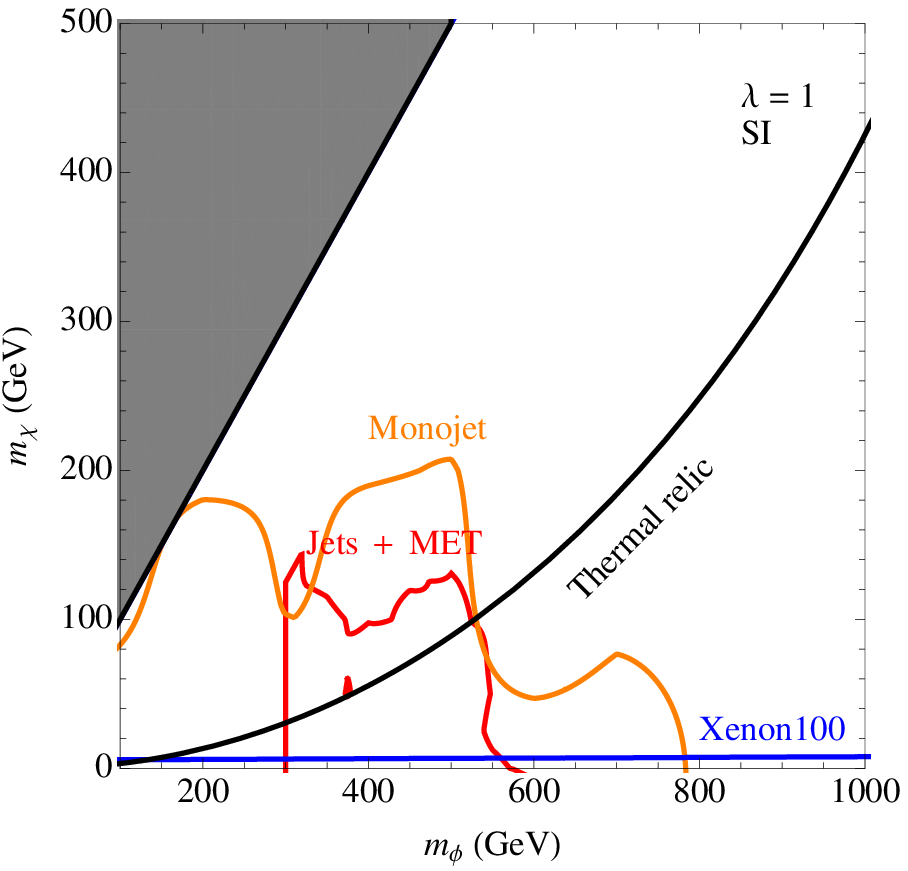} \hspace{3mm}
\includegraphics[width=0.45\textwidth]{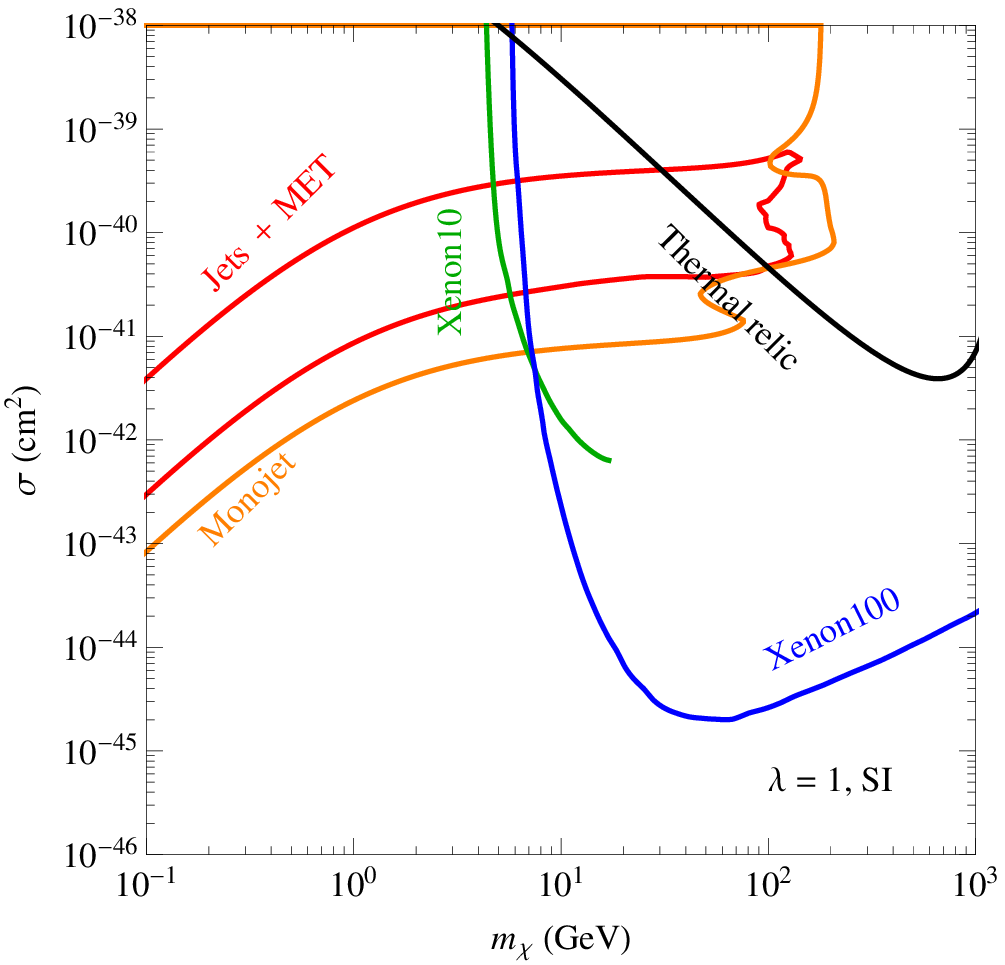}
\caption{95\% exclusion limits from the most sensitive searches for Dirac
  dark matter with coupling to the up quark.}
\label{fig:bounds-dir-l1}
\end{center}
\end{figure}

\begin{figure}[th!]
\begin{center}
\hspace*{-0.75cm}
\includegraphics[width=0.448\textwidth]{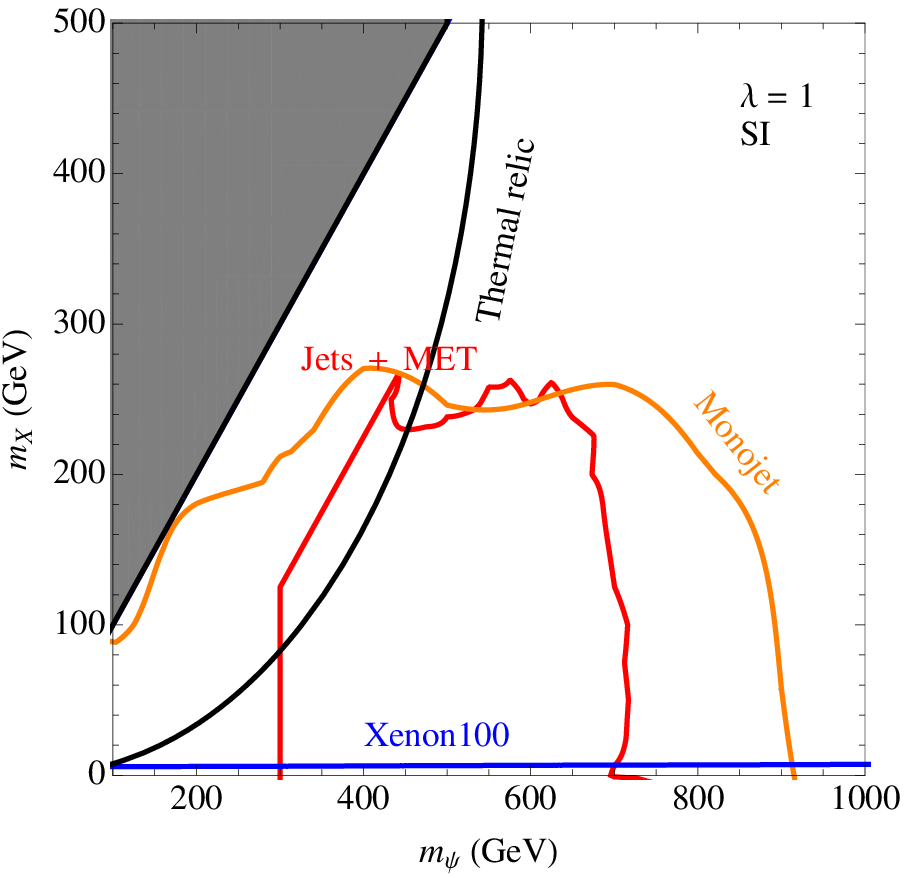} \hspace{3mm}
\includegraphics[width=0.45\textwidth]{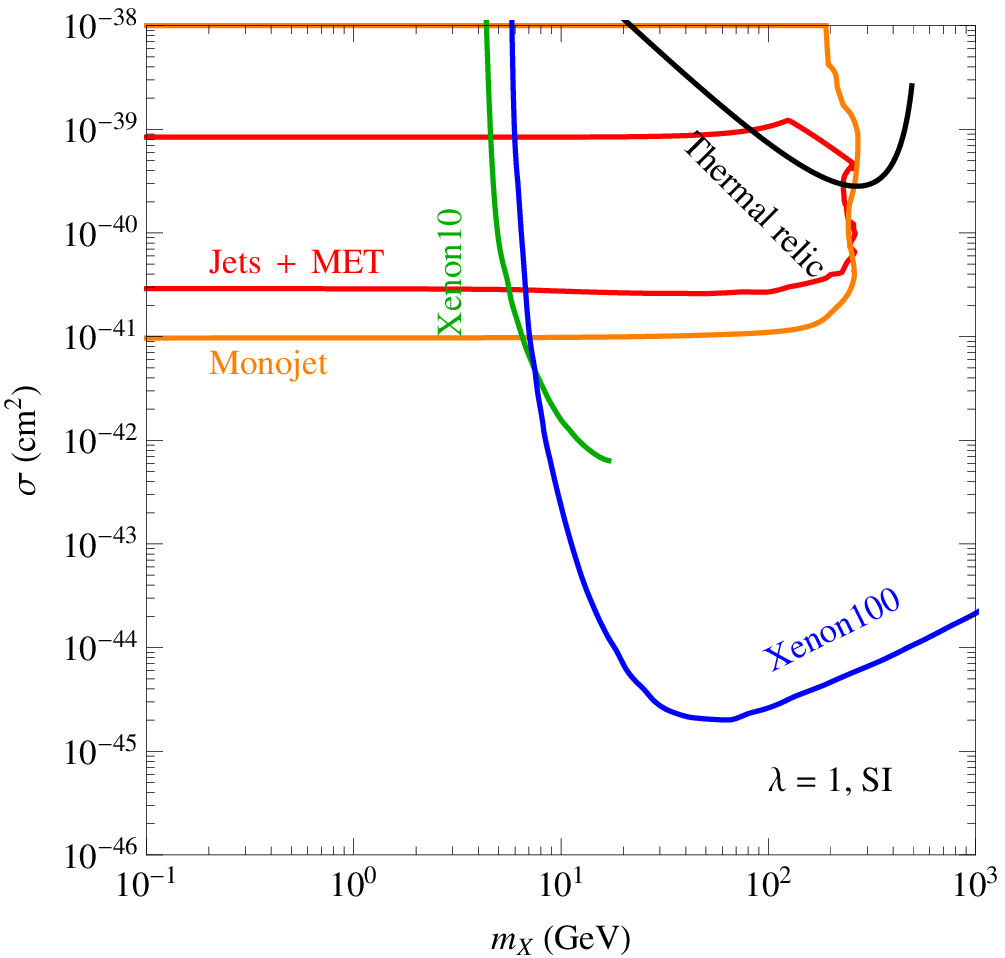}
\caption{95\% exclusion limits from the most sensitive searches for complex scalar
  dark matter with coupling to the up quark.}
\label{fig:bounds-comp-l1}
\end{center}
\end{figure}

Next, we consider models with Dirac dark matter and complex scalar
dark matter. For these models, the SI direct detection 
constraints dominate up to very low dark matter masses, independent of
$m_\phi$.  For $\lambda_{u} = 1$, the exclusion curves are shown in
Figs.~\ref{fig:bounds-dir-l1} and \ref{fig:bounds-comp-l1}.
These cases are highly constrained by searches for spin-independent
scattering, which is unsuppressed.  Since dark matter interactions generally
violate isospin in our models, the different couplings to protons and
neutrons should be taken into account in calculating the bounds.  The
SI cross-section bounds per nucleon are generally calculated under the
assumption of isospin, such that the proton and neutron cross-sections
are the same.  In order to take into account isospin violation, we calculate the
cross-section for interaction with a proton and rescale by
\begin{equation}
  \label{eq:rescale}
  \sigma_{{\rm DM},{\rm nucleon}} = \frac{[f_p Z + f_n (A - Z)]^2}{f_p^2 A^2} \sigma_{{\rm DM},p}\,,
\end{equation}
where $A$ and $Z$ are the mass number and atomic number of the target
nucleus respectively.  The dominant SI bounds come from Xe targets, so
that $A = 131$, neglecting small effects from other comparable or subdominant isotopes,
and $Z = 54$.  All scattering cross sections presented in Figs.~\ref{fig:bounds-dir-l1} and \ref{fig:bounds-comp-l1} are the averaged one, $\sigma_{{\rm DM},{\rm nucleon}}$. 

It is interesting to note that collider bounds take over for light dark
matter, below the threshold of direct detection experiments.  In the
case of a complex scalar, the low mass bound flattens out in the
cross-section plane since it is not sensitive to the reduced mass of
the dark matter-nucleon system, but rather the nucleon mass itself, as can be seen from Eq.~(\ref{eq:si-complex-scalar}).

\section{Discussion and conclusions}
\label{sec:conclusion}
\begin{figure}[th!]
\begin{center}
\hspace*{-0.75cm}
\includegraphics[width=0.45\textwidth]{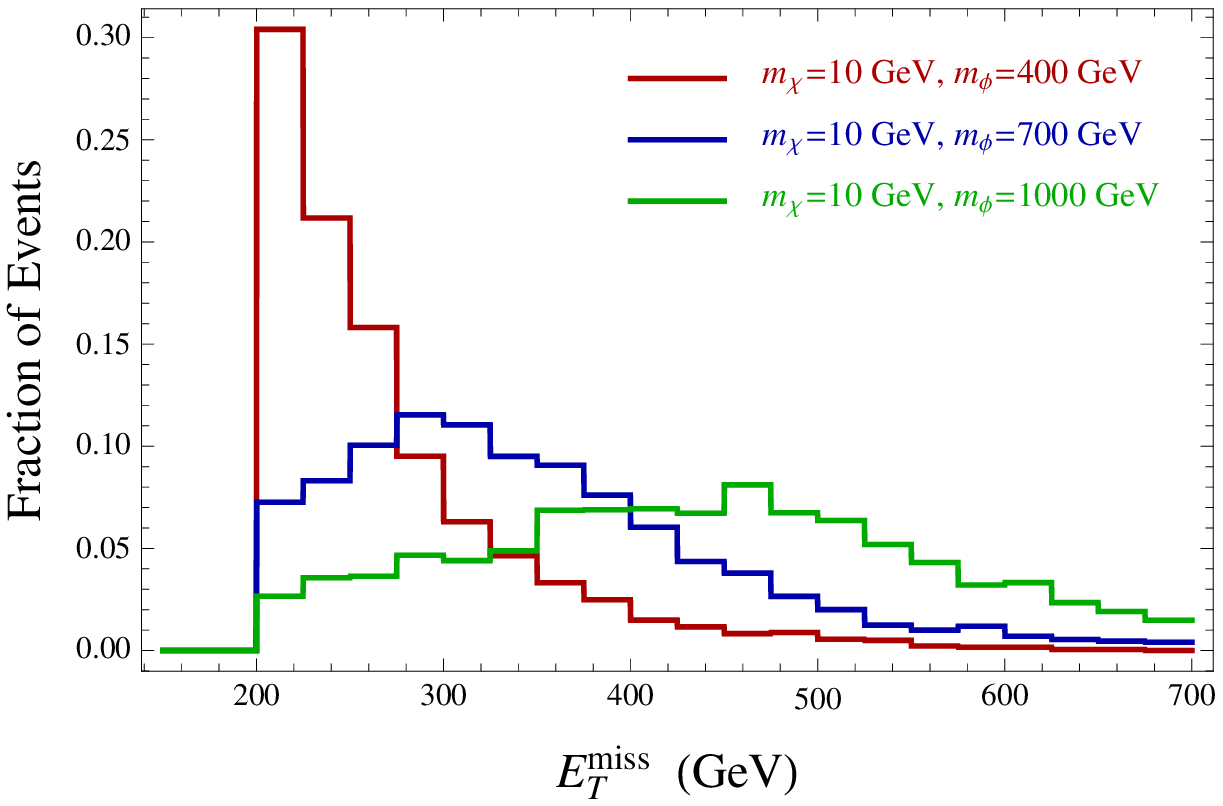} \hspace{3mm}
\includegraphics[width=0.45\textwidth]{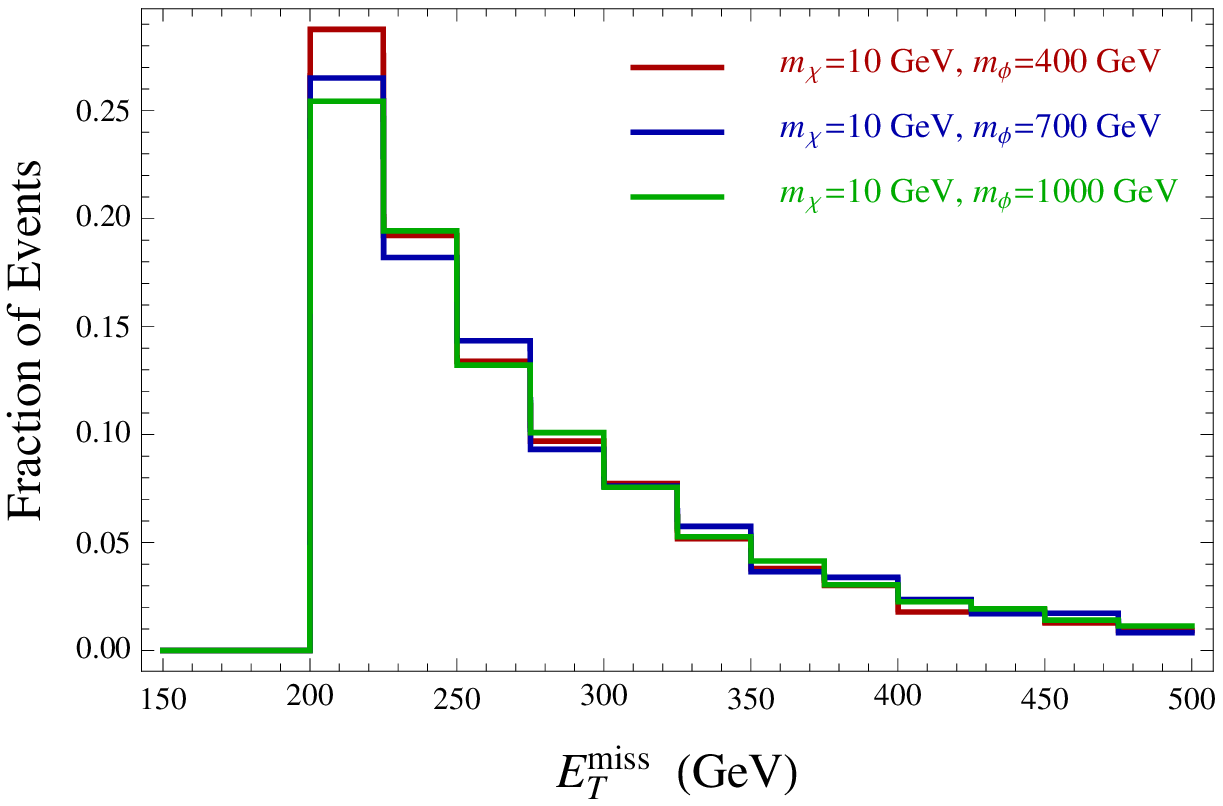}
\caption{Left panel: the fraction of events after basic cuts as a function of $E_T^{\rm miss}$ for the associated production of $\chi + \phi$ with $\phi \rightarrow \chi + j$. Right panel: the same as the left one but for the productions of $2\chi + j$ with the jet from ISR.}
\label{fig:ET-shape}
\end{center}
\end{figure}

The signal spectrum from the associated production of dark matter and
its partner could be dramatically different from
backgrounds. Particularly when the Yukawa coupling is small,
associated production is the dominant part of the signal.  Additional
kinematic variables can be used to enhance the dark matter signal in
the fermion-portal scenario. We use
\texttt{MadGraph5}~\cite{Alwall:2011uj} to generate the dark matter
signal events and shower them in 
\texttt{PYTHIA}~\cite{Sjostrand:2006za}.  We then use
\texttt{PGS}~\cite{PGS} to perform the fast detector simulation. After
utilizing the basic cuts in Ref.~\cite{CMS-PAS-EXO-12-048}, where $E_T^{\rm
  miss} > 200$~GeV has been imposed, we calculate the normalized $E_T^{\rm
  miss}$ distributions for several different spectra.  In the left
panel of Fig.~\ref{fig:ET-shape}, we show the $E_T^{\rm miss}$ from
the $\chi + \phi$ associate productions. Because the jet from the
decay of $\phi \rightarrow \chi + j$ is energetic, the $E_T^{\rm
  miss}$ distributions have a peak-structure with the peak at around
$m_\phi /2$ for a small $m_\chi$.  As a comparison, the right panel of the
Fig.~\ref{fig:ET-shape} shows the $E_T^{\rm miss}$ distribution without
on-shell production of $\phi$.  The spectrum is monotonically
decreasing in this case, which
follows the shape of the background although with a different slope.  For a larger $m_\phi$, the signal
spectrum becomes slightly harder at higher masses.  
In principle, the peak structure in the left panel can be used to
discover dark matter, for instance performing a ``bump" search in the
$E_T^{\rm miss}$ distribution. In practice, the peaks are too wide to
make it feasible. Improving the jet energy resolution and $E_T^{\rm
  miss}$ measurement can yield significant boosts in sensitivity.

To explore more fermion portal dark matter parameter space, we
emphasize the importance of a dedicated search of the two jets plus
MET signature. As can be seen from the left panel in
Fig.~\ref{fig:other-bounds-maj-up}, for small values of the Yukawa
coupling, the current limit on the colored mediator mass is weak,
around 350 GeV for a dark matter mass at 100 GeV. Additional kinematic
variables like $m_{T_2}$ can increase the search
sensitivity~\cite{Barr:2003rg}.

In this paper, we have concentrated on the complimentarity of collider and direct detection searches. The indirect searches for dark matter should also play a role, although the current constraint is not that sensitive. For instance, we have checked that for $\lambda=1$, a dark matter mass of 100 GeV and any mediator mass above the dark matter mass, the new contribution to the anti-proton flux from Dirac fermion dark matter annihilations is at least one order of magnitude below the measured flux at PAMELA~\cite{Adriani:2010rc}. Because of the $p$-wave suppression for the Majorana fermion and complex scalar dark matter annihilation cases, the indirect detection constraints are even weaker.

\vspace{3mm}
Note added: we note here that during the completion of our paper, another paper~\cite{Chang:2013oia} appeared with different emphasis: our paper concentrates more on the complimentarity of direct detection and collider searches for dark matter, while their paper has more focus on the thermal relic parameter space.

\subsection*{Acknowledgments} 
Y. Bai is supported by startup funds from the UW-Madison. SLAC is operated by Stanford University for the US Department of Energy under contract DE-AC02-76SF00515. YB also would like to thank the Kvali Institute for Theoretical Physics, U. C. Santa Barbara, where part of this work was done. This research was also supported in part by the National Science Foundation under Grant No. NSF PHY11-25915.


\begin{thebibliography}{10}

\bibitem{Begeman:1991iy}
K.~Begeman, A.~Broeils, and R.~Sanders, {\it {Extended rotation curves of
  spiral galaxies: Dark haloes and modified dynamics}},  {\em
  Mon.Not.Roy.Astron.Soc.} {\bf 249} (1991) 523.

\bibitem{Bradac:2006er}
M.~Bradac, D.~Clowe, A.~H. Gonzalez, P.~Marshall, W.~Forman, {\em et.~al.},
  {\it {Strong and weak lensing united. 3. Measuring the mass distribution of
  the merging galaxy cluster 1E0657-56}},  {\em Astrophys.J.} {\bf 652} (2006)
  937--947, [\href{http://xxx.lanl.gov/abs/astro-ph/0608408}{{\tt
  astro-ph/0608408}}].

\bibitem{Bennett:2012zja}
{\bf WMAP} Collaboration, C.~Bennett {\em et.~al.}, {\it {Nine-Year Wilkinson
  Microwave Anisotropy Probe (WMAP) Observations: Final Maps and Results}},
  \href{http://xxx.lanl.gov/abs/1212.5225}{{\tt arXiv:1212.5225}}.

\bibitem{Ade:2013zuv}
{\bf Planck} Collaboration, P.~Ade {\em et.~al.}, {\it {Planck 2013 results.
  XVI. Cosmological parameters}},
  \href{http://xxx.lanl.gov/abs/1303.5076}{{\tt arXiv:1303.5076}}.

\bibitem{Birkedal:2004xn}
A.~Birkedal, K.~Matchev, and M.~Perelstein, {\it {Dark matter at colliders: A
  Model independent approach}},  {\em Phys.Rev.} {\bf D70} (2004) 077701,
  [\href{http://xxx.lanl.gov/abs/hep-ph/0403004}{{\tt hep-ph/0403004}}].

\bibitem{Barger:2008qd}
V.~Barger, W.-Y. Keung, and G.~Shaughnessy, {\it {Spin Dependence of Dark
  Matter Scattering}},  {\em Phys.Rev.} {\bf D78} (2008) 056007,
  [\href{http://xxx.lanl.gov/abs/0806.1962}{{\tt arXiv:0806.1962}}].

\bibitem{Cao:2009uw}
Q.-H. Cao, C.-R. Chen, C.~S. Li, and H.~Zhang, {\it {Effective Dark Matter
  Model: Relic density, CDMS II, Fermi LAT and LHC}},  {\em JHEP} {\bf 1108}
  (2011) 018, [\href{http://xxx.lanl.gov/abs/0912.4511}{{\tt
  arXiv:0912.4511}}].

\bibitem{Beltran:2010ww}
M.~Beltran, D.~Hooper, E.~W. Kolb, Z.~A. Krusberg, and T.~M. Tait, {\it
  {Maverick dark matter at colliders}},  {\em JHEP} {\bf 1009} (2010) 037,
  [\href{http://xxx.lanl.gov/abs/1002.4137}{{\tt arXiv:1002.4137}}].

\bibitem{Bai:2010hh}
Y.~Bai, P.~J. Fox, and R.~Harnik, {\it {The Tevatron at the Frontier of Dark
  Matter Direct Detection}},  {\em JHEP} {\bf 1012} (2010) 048,
  [\href{http://xxx.lanl.gov/abs/1005.3797}{{\tt arXiv:1005.3797}}].

\bibitem{Goodman:2010yf}
J.~Goodman, M.~Ibe, A.~Rajaraman, W.~Shepherd, T.~M. Tait, {\em et.~al.}, {\it
  {Constraints on Light Majorana dark Matter from Colliders}},  {\em
  Phys.Lett.} {\bf B695} (2011) 185--188,
  [\href{http://xxx.lanl.gov/abs/1005.1286}{{\tt arXiv:1005.1286}}].

\bibitem{Goodman:2010ku}
J.~Goodman, M.~Ibe, A.~Rajaraman, W.~Shepherd, T.~M. Tait, {\em et.~al.}, {\it
  {Constraints on Dark Matter from Colliders}},  {\em Phys.Rev.} {\bf D82}
  (2010) 116010, [\href{http://xxx.lanl.gov/abs/1008.1783}{{\tt
  arXiv:1008.1783}}].

\bibitem{Fan:2010gt}
J.~Fan, M.~Reece, and L.-T. Wang, {\it {Non-relativistic effective theory of
  dark matter direct detection}},  {\em JCAP} {\bf 1011} (2010) 042,
  [\href{http://xxx.lanl.gov/abs/1008.1591}{{\tt arXiv:1008.1591}}].

\bibitem{Buckley:2011kk}
M.~R. Buckley, {\it {Asymmetric Dark Matter and Effective Operators}},  {\em
  Phys.Rev.} {\bf D84} (2011) 043510,
  [\href{http://xxx.lanl.gov/abs/1104.1429}{{\tt arXiv:1104.1429}}].

\bibitem{Rajaraman:2011wf}
A.~Rajaraman, W.~Shepherd, T.~M. Tait, and A.~M. Wijangco, {\it {LHC Bounds on
  Interactions of Dark Matter}},  {\em Phys.Rev.} {\bf D84} (2011) 095013,
  [\href{http://xxx.lanl.gov/abs/1108.1196}{{\tt arXiv:1108.1196}}].

\bibitem{Fox:2011pm}
P.~J. Fox, R.~Harnik, J.~Kopp, and Y.~Tsai, {\it {Missing Energy Signatures of
  Dark Matter at the LHC}},  {\em Phys.Rev.} {\bf D85} (2012) 056011,
  [\href{http://xxx.lanl.gov/abs/1109.4398}{{\tt arXiv:1109.4398}}].

\bibitem{Cheung:2012gi}
K.~Cheung, P.-Y. Tseng, Y.-L.~S. Tsai, and T.-C. Yuan, {\it {Global Constraints
  on Effective Dark Matter Interactions: Relic Density, Direct Detection,
  Indirect Detection, and Collider}},  {\em JCAP} {\bf 1205} (2012) 001,
  [\href{http://xxx.lanl.gov/abs/1201.3402}{{\tt arXiv:1201.3402}}].

\bibitem{Aaltonen:2012jb}
{\bf CDF} Collaboration, T.~Aaltonen {\em et.~al.}, {\it {A Search for dark
  matter in events with one jet and missing transverse energy in $p\bar{p}$
  collisions at $\sqrt{s} = 1.96$ TeV}},  {\em Phys.Rev.Lett.} {\bf 108} (2012)
  211804, [\href{http://xxx.lanl.gov/abs/1203.0742}{{\tt arXiv:1203.0742}}].

\bibitem{Fitzpatrick:2012ix}
A.~L. Fitzpatrick, W.~Haxton, E.~Katz, N.~Lubbers, and Y.~Xu, {\it {The
  Effective Field Theory of Dark Matter Direct Detection}},  {\em JCAP} {\bf
  1302} (2013) 004, [\href{http://xxx.lanl.gov/abs/1203.3542}{{\tt
  arXiv:1203.3542}}].

\bibitem{Barger:2012pf}
V.~Barger, W.-Y. Keung, D.~Marfatia, and P.-Y. Tseng, {\it {Dipole Moment Dark
  Matter at the LHC}},  {\em Phys.Lett.} {\bf B717} (2012) 219--223,
  [\href{http://xxx.lanl.gov/abs/1206.0640}{{\tt arXiv:1206.0640}}].

\bibitem{Bai:2012xg}
Y.~Bai and T.~M. Tait, {\it {Searches with Mono-Leptons}},  {\em Phys.Lett.}
  {\bf B723} (2013) 384--387, [\href{http://xxx.lanl.gov/abs/1208.4361}{{\tt
  arXiv:1208.4361}}].

\bibitem{ATLAS-CONF-2012-147}
{\it Search for new phenomena in monojet plus missing transverse momentum final
  states using 10fb-1 of pp collisions at sqrt{s}=8 tev with the atlas detector
  at the lhc},  Tech. Rep. ATLAS-CONF-2012-147, CERN, Geneva, Nov, 2012.

\bibitem{Chae:2012bq}
Y.~J. Chae and M.~Perelstein, {\it {Dark Matter Search at a Linear Collider:
  Effective Operator Approach}},  {\em JHEP} {\bf 1305} (2013) 138,
  [\href{http://xxx.lanl.gov/abs/1211.4008}{{\tt arXiv:1211.4008}}].

\bibitem{Dreiner:2013vla}
H.~Dreiner, D.~Schmeier, and J.~Tattersall, {\it {Contact Interactions Probe
  Effective Dark Matter Models at the LHC}},  {\em Europhys.Lett.} {\bf 102}
  (2013) 51001, [\href{http://xxx.lanl.gov/abs/1303.3348}{{\tt
  arXiv:1303.3348}}].

\bibitem{Cornell:2013rza}
J.~M. Cornell, S.~Profumo, and W.~Shepherd, {\it {Kinetic Decoupling and
  Small-Scale Structure in Effective Theories of Dark Matter}},
  \href{http://xxx.lanl.gov/abs/1305.4676}{{\tt arXiv:1305.4676}}.

\bibitem{CMS-PAS-EXO-12-048}
{\it Search for new physics in monojet events in pp collisions at sqrt(s)= 8
  tev},  Tech. Rep. CMS-PAS-EXO-12-048, CERN, Geneva, 2013.

\bibitem{Shoemaker:2011vi}
I.~M. Shoemaker and L.~Vecchi, {\it {Unitarity and Monojet Bounds on Models for
  DAMA, CoGeNT, and CRESST-II}},  {\em Phys.Rev.} {\bf D86} (2012) 015023,
  [\href{http://xxx.lanl.gov/abs/1112.5457}{{\tt arXiv:1112.5457}}].

\bibitem{An:2012ue}
H.~An, R.~Huo, and L.-T. Wang, {\it {Searching for Low Mass Dark Portal at the
  LHC}},  {\em Phys.Dark Univ.} {\bf 2} (2013) 50--57,
  [\href{http://xxx.lanl.gov/abs/1212.2221}{{\tt arXiv:1212.2221}}].

\bibitem{Busoni:2013lha}
G.~Busoni, A.~De~Simone, E.~Morgante, and A.~Riotto, {\it {On the Validity of
  the Effective Field Theory for Dark Matter Searches at the LHC}},
  \href{http://xxx.lanl.gov/abs/1307.2253}{{\tt arXiv:1307.2253}}.

\bibitem{ATLAS-CONF-2013-073}
{\it Search for dark matter pair production in events with a hadronically
  decaying w or z boson and missing transverse momentum in pp collision data at
  sqrt(s) = 8 tev with the atlas detector},  Tech. Rep. ATLAS-CONF-2013-073,
  CERN, Geneva, July, 2013.

\bibitem{CMS-PAS-EXO-13-004}
{\it Mono-lepton dark matter interpretation at sqrt{s} = 8 tev},  Tech. Rep.
  CMS-PAS-EXO-13-004, CERN, Geneva, 2013.

\bibitem{Cahill-Rowley:2013dpa}
M.~Cahill-Rowley, R.~Cotta, A.~Drlica-Wagner, S.~Funk, J.~Hewett, {\em
  et.~al.}, {\it {Complementarity and Searches for Dark Matter in the pMSSM}},
  \href{http://xxx.lanl.gov/abs/1305.6921}{{\tt arXiv:1305.6921}}.

\bibitem{Shrock:1982kd}
R.~E. Shrock and M.~Suzuki, {\it {INVISIBLE DECAYS OF HIGGS BOSONS}},  {\em
  Phys.Lett.} {\bf B110} (1982) 250.

\bibitem{Burgess:2000yq}
C.~Burgess, M.~Pospelov, and T.~ter Veldhuis, {\it {The Minimal model of
  nonbaryonic dark matter: A Singlet scalar}},  {\em Nucl.Phys.} {\bf B619}
  (2001) 709--728, [\href{http://xxx.lanl.gov/abs/hep-ph/0011335}{{\tt
  hep-ph/0011335}}].

\bibitem{Patt:2006fw}
B.~Patt and F.~Wilczek, {\it {Higgs-field portal into hidden sectors}},
  \href{http://xxx.lanl.gov/abs/hep-ph/0605188}{{\tt hep-ph/0605188}}.

\bibitem{Bai:2012nv}
Y.~Bai, V.~Barger, L.~L. Everett, and G.~Shaughnessy, {\it {2HDM Portal Dark
  Matter: LHC data and the Fermi-LAT 135 GeV Line}},
  \href{http://xxx.lanl.gov/abs/1212.5604}{{\tt arXiv:1212.5604}}.

\bibitem{Nomura:2008ru}
Y.~Nomura and J.~Thaler, {\it {Dark Matter through the Axion Portal}},  {\em
  Phys.Rev.} {\bf D79} (2009) 075008,
  [\href{http://xxx.lanl.gov/abs/0810.5397}{{\tt arXiv:0810.5397}}].

\bibitem{Lee:2013bua}
H.~M. Lee, M.~Park, and V.~Sanz, {\it {Gravity-mediated Dark Matter}},
  \href{http://xxx.lanl.gov/abs/1306.4107}{{\tt arXiv:1306.4107}}.

\bibitem{Bai:2009ms}
Y.~Bai, M.~Carena, and J.~Lykken, {\it {Dilaton-assisted Dark Matter}},  {\em
  Phys.Rev.Lett.} {\bf 103} (2009) 261803,
  [\href{http://xxx.lanl.gov/abs/0909.1319}{{\tt arXiv:0909.1319}}].

\bibitem{Cheung:2007ut}
K.~Cheung and T.-C. Yuan, {\it {Hidden fermion as milli-charged dark matter in
  Stueckelberg Z- prime model}},  {\em JHEP} {\bf 0703} (2007) 120,
  [\href{http://xxx.lanl.gov/abs/hep-ph/0701107}{{\tt hep-ph/0701107}}].

\bibitem{Feldman:2007wj}
D.~Feldman, Z.~Liu, and P.~Nath, {\it {The Stueckelberg Z-prime Extension with
  Kinetic Mixing and Milli-Charged Dark Matter From the Hidden Sector}},  {\em
  Phys.Rev.} {\bf D75} (2007) 115001,
  [\href{http://xxx.lanl.gov/abs/hep-ph/0702123}{{\tt hep-ph/0702123}}].

\bibitem{Shiu:2013wxa}
G.~Shiu, P.~Soler, and F.~Ye, {\it {Millicharged Dark Matter in Quantum Gravity
  and String Theory}},  {\em Phys. Rev. Lett.} {\bf 110} (2013) 241304,
  [\href{http://xxx.lanl.gov/abs/1302.5471}{{\tt arXiv:1302.5471}}].

\bibitem{Banks:2010zn}
T.~Banks and N.~Seiberg, {\it {Symmetries and Strings in Field Theory and
  Gravity}},  {\em Phys.Rev.} {\bf D83} (2011) 084019,
  [\href{http://xxx.lanl.gov/abs/1011.5120}{{\tt arXiv:1011.5120}}].

\bibitem{Dudas:2009uq}
E.~Dudas, Y.~Mambrini, S.~Pokorski, and A.~Romagnoni, {\it {(In)visible Z-prime
  and dark matter}},  {\em JHEP} {\bf 0908} (2009) 014,
  [\href{http://xxx.lanl.gov/abs/0904.1745}{{\tt arXiv:0904.1745}}].

\bibitem{Jungman:1995df}
G.~Jungman, M.~Kamionkowski, and K.~Griest, {\it {Supersymmetric dark matter}},
   {\em Phys.Rept.} {\bf 267} (1996) 195--373,
  [\href{http://xxx.lanl.gov/abs/hep-ph/9506380}{{\tt hep-ph/9506380}}].

\bibitem{Bertone:2004pz}
G.~Bertone, D.~Hooper, and J.~Silk, {\it {Particle dark matter: Evidence,
  candidates and constraints}},  {\em Phys.Rept.} {\bf 405} (2005) 279--390,
  [\href{http://xxx.lanl.gov/abs/hep-ph/0404175}{{\tt hep-ph/0404175}}].

\bibitem{Agrawal:2010fh}
P.~Agrawal, Z.~Chacko, C.~Kilic, and R.~K. Mishra, {\it {A Classification of
  Dark Matter Candidates with Primarily Spin-Dependent Interactions with
  Matter}},  \href{http://xxx.lanl.gov/abs/1003.1912}{{\tt arXiv:1003.1912}}.

\bibitem{Garny:2013ama}
M.~Garny, A.~Ibarra, M.~Pato, and S.~Vogl, {\it {Internal bremsstrahlung
  signatures in light of direct dark matter searches}},
  \href{http://xxx.lanl.gov/abs/1306.6342}{{\tt arXiv:1306.6342}}.

\bibitem{Griest:1990kh}
K.~Griest and D.~Seckel, {\it {Three exceptions in the calculation of relic
  abundances}},  {\em Phys.Rev.} {\bf D43} (1991) 3191--3203.

\bibitem{Edsjo:1997bg}
J.~Edsjo and P.~Gondolo, {\it {Neutralino relic density including
  coannihilations}},  {\em Phys.Rev.} {\bf D56} (1997) 1879--1894,
  [\href{http://xxx.lanl.gov/abs/hep-ph/9704361}{{\tt hep-ph/9704361}}].

\bibitem{Gondolo:2004sc}
P.~Gondolo, J.~Edsjo, P.~Ullio, L.~Bergstrom, M.~Schelke, {\em et.~al.}, {\it
  {DarkSUSY: Computing supersymmetric dark matter properties numerically}},
  {\em JCAP} {\bf 0407} (2004) 008,
  [\href{http://xxx.lanl.gov/abs/astro-ph/0406204}{{\tt astro-ph/0406204}}].

\bibitem{Belanger:2008sj}
G.~Belanger, F.~Boudjema, A.~Pukhov, and A.~Semenov, {\it {Dark matter direct
  detection rate in a generic model with micrOMEGAs 2.2}},  {\em
  Comput.Phys.Commun.} {\bf 180} (2009) 747--767,
  [\href{http://xxx.lanl.gov/abs/0803.2360}{{\tt arXiv:0803.2360}}].

\bibitem{Aprile:2012nq}
{\bf XENON100} Collaboration, E.~Aprile {\em et.~al.}, {\it {Dark Matter
  Results from 225 Live Days of XENON100 Data}},  {\em Phys.Rev.Lett.} {\bf
  109} (2012) 181301, [\href{http://xxx.lanl.gov/abs/1207.5988}{{\tt
  arXiv:1207.5988}}].

\bibitem{Angle:2011th}
{\bf XENON10} Collaboration, J.~Angle {\em et.~al.}, {\it {A search for light
  dark matter in XENON10 data}},  {\em Phys.Rev.Lett.} {\bf 107} (2011) 051301,
  [\href{http://xxx.lanl.gov/abs/1104.3088}{{\tt arXiv:1104.3088}}].

\bibitem{Felizardo:2011uw}
M.~Felizardo, T.~Girard, T.~Morlat, A.~Fernandes, A.~Ramos, {\em et.~al.}, {\it
  {Final Analysis and Results of the Phase II SIMPLE Dark Matter Search}},
  {\em Phys.Rev.Lett.} {\bf 108} (2012) 201302,
  [\href{http://xxx.lanl.gov/abs/1106.3014}{{\tt arXiv:1106.3014}}].

\bibitem{Behnke:2010xt}
E.~Behnke, J.~Behnke, S.~Brice, D.~Broemmelsiek, J.~Collar, {\em et.~al.}, {\it
  {Improved Limits on Spin-Dependent WIMP-Proton Interactions from a Two Liter
  CF$_3$I Bubble Chamber}},  {\em Phys.Rev.Lett.} {\bf 106} (2011) 021303,
  [\href{http://xxx.lanl.gov/abs/1008.3518}{{\tt arXiv:1008.3518}}].

\bibitem{Archambault:2009sm}
S.~Archambault, F.~Aubin, M.~Auger, E.~Behnke, B.~Beltran, {\em et.~al.}, {\it
  {Dark Matter Spin-Dependent Limits for WIMP Interactions on F-19 by
  PICASSO}},  {\em Phys.Lett.} {\bf B682} (2009) 185--192,
  [\href{http://xxx.lanl.gov/abs/0907.0307}{{\tt arXiv:0907.0307}}].

\bibitem{Aprile:2013doa}
{\bf XENON100} Collaboration, E.~Aprile {\em et.~al.}, {\it {Limits on
  spin-dependent WIMP-nucleon cross sections from 225 live days of XENON100
  data}},  \href{http://xxx.lanl.gov/abs/1301.6620}{{\tt arXiv:1301.6620}}.

\bibitem{Ahmed:2008eu}
{\bf CDMS} Collaboration, Z.~Ahmed {\em et.~al.}, {\it {Search for Weakly
  Interacting Massive Particles with the First Five-Tower Data from the
  Cryogenic Dark Matter Search at the Soudan Underground Laboratory}},  {\em
  Phys.Rev.Lett.} {\bf 102} (2009) 011301,
  [\href{http://xxx.lanl.gov/abs/0802.3530}{{\tt arXiv:0802.3530}}].

\bibitem{Ahmed:2010wy}
{\bf CDMS-II} Collaboration, Z.~Ahmed {\em et.~al.}, {\it {Results from a
  Low-Energy Analysis of the CDMS II Germanium Data}},  {\em Phys.Rev.Lett.}
  {\bf 106} (2011) 131302, [\href{http://xxx.lanl.gov/abs/1011.2482}{{\tt
  arXiv:1011.2482}}].

\bibitem{Bernabei:2013cfa}
R.~Bernabei, P.~Belli, S.~d'Angelo, A.~Di~Marco, F.~Montecchia, {\em et.~al.},
  {\it {Dark Matter investigation by DAMA at Gran Sasso}},  {\em
  Int.J.Mod.Phys.} {\bf A28} (2013) 1330022,
  [\href{http://xxx.lanl.gov/abs/1306.1411}{{\tt arXiv:1306.1411}}].

\bibitem{Aalseth:2010vx}
{\bf CoGeNT} Collaboration, C.~Aalseth {\em et.~al.}, {\it {Results from a
  Search for Light-Mass Dark Matter with a P-type Point Contact Germanium
  Detector}},  {\em Phys.Rev.Lett.} {\bf 106} (2011) 131301,
  [\href{http://xxx.lanl.gov/abs/1002.4703}{{\tt arXiv:1002.4703}}].

\bibitem{Angloher:2011uu}
G.~Angloher, M.~Bauer, I.~Bavykina, A.~Bento, C.~Bucci, {\em et.~al.}, {\it
  {Results from 730 kg days of the CRESST-II Dark Matter Search}},  {\em
  Eur.Phys.J.} {\bf C72} (2012) 1971,
  [\href{http://xxx.lanl.gov/abs/1109.0702}{{\tt arXiv:1109.0702}}].

\bibitem{Agnese:2013rvf}
{\bf CDMS} Collaboration, R.~Agnese {\em et.~al.}, {\it {Dark Matter Search
  Results Using the Silicon Detectors of CDMS II}},  {\em Phys.Rev.Lett.}
  (2013) [\href{http://xxx.lanl.gov/abs/1304.4279}{{\tt arXiv:1304.4279}}].

\bibitem{Feng:2013vod}
J.~L. Feng, J.~Kumar, and D.~Sanford, {\it {Xenophobic Dark Matter}},
  \href{http://xxx.lanl.gov/abs/1306.2315}{{\tt arXiv:1306.2315}}.

\bibitem{CMS-PAS-SUS-13-012}
{\it Search for new physics in the multijets and missing momentum final state
  in proton-proton collisions at 8 tev},  Tech. Rep. CMS-PAS-SUS-13-012, CERN,
  Geneva, 2013.

\bibitem{ATLAS-CONF-2013-047}
{\it Search for squarks and gluinos with the atlas detector in final states
  with jets and missing transverse momentum and 20.3 fb$^{-1}$ of $\sqrt{s}=8$
  tev proton-proton collision data},  Tech. Rep. ATLAS-CONF-2013-047, CERN,
  Geneva, May, 2013.

\bibitem{Chatrchyan:2013lya}
{\bf CMS} Collaboration, S.~Chatrchyan {\em et.~al.}, {\it {Search for
  supersymmetry in hadronic final states with missing transverse energy using
  the variables $\alpha_T$ and b-quark multiplicity in pp collisions at
  $\sqrt{s}$ = 8 TeV}},  \href{http://xxx.lanl.gov/abs/1303.2985}{{\tt
  arXiv:1303.2985}}.

\bibitem{Alwall:2011uj}
J.~Alwall, M.~Herquet, F.~Maltoni, O.~Mattelaer, and T.~Stelzer, {\it {MadGraph
  5 : Going Beyond}},  {\em JHEP} {\bf 1106} (2011) 128,
  [\href{http://xxx.lanl.gov/abs/1106.0522}{{\tt arXiv:1106.0522}}].

\bibitem{Christensen:2008py}
N.~D. Christensen and C.~Duhr, {\it {FeynRules - Feynman rules made easy}},
  {\em Comput.Phys.Commun.} {\bf 180} (2009) 1614--1641,
  [\href{http://xxx.lanl.gov/abs/0806.4194}{{\tt arXiv:0806.4194}}].

\bibitem{Beenakker:1996ed}
W.~Beenakker, R.~Hopker, and M.~Spira, {\it {PROSPINO: A Program for the
  production of supersymmetric particles in next-to-leading order QCD}},
  \href{http://xxx.lanl.gov/abs/hep-ph/9611232}{{\tt hep-ph/9611232}}.

\bibitem{Sjostrand:2007gs}
T.~Sjostrand, S.~Mrenna, and P.~Z. Skands, {\it {A Brief Introduction to PYTHIA
  8.1}},  {\em Comput.Phys.Commun.} {\bf 178} (2008) 852--867,
  [\href{http://xxx.lanl.gov/abs/0710.3820}{{\tt arXiv:0710.3820}}].

\bibitem{Cacciari:2011ma}
M.~Cacciari, G.~P. Salam, and G.~Soyez, {\it {FastJet User Manual}},  {\em
  Eur.Phys.J.} {\bf C72} (2012) 1896,
  [\href{http://xxx.lanl.gov/abs/1111.6097}{{\tt arXiv:1111.6097}}].

\bibitem{Sjostrand:2006za}
T.~Sjostrand, S.~Mrenna, and P.~Z. Skands, {\it {PYTHIA 6.4 Physics and
  Manual}},  {\em JHEP} {\bf 0605} (2006) 026,
  [\href{http://xxx.lanl.gov/abs/hep-ph/0603175}{{\tt hep-ph/0603175}}].

\bibitem{PGS}
J.~S. Conway, {\it {Pretty Good Simulation of high-energy collisions}},
  \href{http://xxx.lanl.gov/abs/090401 release}{{\tt 090401 release}}.

\bibitem{Barr:2003rg}
A.~Barr, C.~Lester, and P.~Stephens, {\it {m(T2): The Truth behind the
  glamour}},  {\em J.Phys.} {\bf G29} (2003) 2343--2363,
  [\href{http://xxx.lanl.gov/abs/hep-ph/0304226}{{\tt hep-ph/0304226}}].
  
  \bibitem{Adriani:2010rc}
{\bf PAMELA} Collaboration, O.~Adriani {\em et.~al.}, {\it {PAMELA results on the cosmic-ray antiproton flux from 60 MeV to 180 GeV in kinetic energy}},  {\em Phys.Rev.Lett.}
  (2010) [\href{http://xxx.lanl.gov/abs/1007.0821}{{\tt arXiv:1007.0821}}].

\bibitem{Chang:2013oia}
S.~Chang, R.~Edezhath, J.~Hutchinson, and M.~Luty, {\it {Effective WIMPs}},
  \href{http://xxx.lanl.gov/abs/1307.8120}{{\tt arXiv:1307.8120}}.



\end{thebibliography}
\providecommand{\href}[2]{#2}\begingroup\raggedright\endgroup

 \end{document}